\documentclass[aps,prd,twocolumn,notitlepage,nofootinbib,floatfix,superscriptaddress]{revtex4-1}
\usepackage{amsmath, amsthm, amssymb, amsfonts, amsbsy,mathrsfs}
\usepackage{graphicx}
\usepackage{calligra,bm}
\usepackage{stmaryrd}
\usepackage{xcolor}
\usepackage[hidelinks,bookmarks=true]{hyperref}
\hypersetup{pdfstartview=FitH,pdfhighlight=/O,colorlinks=false}
\graphicspath{{./images/}}

\begin{document}

\title{New class of generalized coupling theories}
\author{Justin C. Feng}
\affiliation{CENTRA, Departamento de F{\'i}sica, Instituto Superior T{\'e}cnico – IST, Universidade de Lisboa – UL, Avenida Rovisco Pais 1, 1049 Lisboa, Portugal}
\author{Sante Carloni}
\affiliation{CENTRA, Departamento de F{\'i}sica, Instituto Superior T{\'e}cnico – IST, Universidade de Lisboa – UL, Avenida Rovisco Pais 1, 1049 Lisboa, Portugal}

%
%
\begin{abstract}
We propose a new class of gravity theories which are characterized by a nontrivial coupling between the gravitational metric and matter mediated by an auxiliary rank-2 tensor. The actions generating the field equations are constructed so that these theories are equivalent to general relativity in a vacuum, and only differ from general relativity theory within a matter distribution. We analyze in detail one of the simplest realizations of these generalized coupling  theories. We show that in this case the propagation speed of gravitational radiation in matter is different from its value in vacuum and that this can be used to weakly constrain the (single) additional parameter of the theory. An analysis of the evolution of homogeneous and isotropic spacetimes in the same framework shows that there exist cosmic histories with both an inflationary phase and a dark era characterized by a different expansion rate.
\end{abstract}
%
%


\maketitle


%
%
\section{Introduction}
In recent years, we have witnessed considerable advances in the accuracy and methodology of experimental and observational investigation of gravitational phenomena. The wealth of new data increasingly exacerbates a puzzle that has been present for several decades, with regard to our current understanding of the gravitational interaction. On one hand, the detection of gravitational waves (e.g., Refs. \citep{LIGOVIRGO_2017MMO,LIGOVIRGO_2015}) and the observation of the black hole at the center of M87 \citep{1435168} have brought extraordinary confirmation of the predictions of general relativity (GR) in the strong field regime. On the other hand, it has become increasingly clear that GR alone is unable to correctly describe the dynamics of objects at galactic and extra-galactic scales \citep{Bertone:2004pz}, the current accelerated expansion of the Universe \citep{Peebles:2002gy}, and the tension in the estimation of the present value of the Hubble parameter \citep{Freedman:2017yms}.

As pointed out in Ref. \citep{Carloni2017}, one way to interpolate between these contrasting results is to reevaluate the interaction between spacetime and matter, rather than assuming that gravity behaves differently at different scales. The motivation for such a point of view lies in the realization that deviations from GR only appear in spacetimes in which the role of matter cannot be neglected, like cosmology and the gravitational behavior of galaxies and clusters of galaxies.

Indeed, the weakest assumption in the construction of the celebrated Einstein equations is the way in which matter and spacetime are coupled to each other. A key principle that guided Einstein was the local conservation of the energy-momentum tensor (the divergence-free property), which in the modern framework is encoded in the existence of a natural variational principle able to generate the field equations \citep{Pyenson1997-PYETCP-2}. However, there is no compelling reason not to consider more complicated connections between the spacetime geometry (Einstein tensor) and the energy-momentum tensor for matter.

If one is willing to consider the possibility that the coupling between these objects is more complex than a simple proportionality, one could consider the following equation \citep{Carloni2017},
\begin{equation}\label{GCA-EFEgen}
G_{\mu \nu} = \chi_{\mu \nu}{^{\alpha \beta}} \, T_{\alpha \beta},
\end{equation}
where the coupling tensor $\chi_{\mu \nu}{^{\alpha \beta}}$ is a generic, nonsingular, fourth-order tensor which mediates (and generalizes) the response of spacetime to a given matter distribution.

The structure of \eqref{GCA-EFEgen} can be engineered in such a way that its phenomenology in vacuum is exactly that of GR. Such a generalization avoids the difficulties that normally afflict modifications of GR. In particular, many modified gravity theories have a nontrivial vacuum phenomenology which is strongly constrained by the measurement of post-Newtonian effects and, more recently, gravitational wave detections and black hole phenomenology. Equation \eqref{GCA-EFEgen} is compatible with all these constraints. Phenomenological differences only appear within a matter distribution, like in the (very different) case of torsion in the Einstein-Cartan-Sciama-Kibble theory \citep{Hehl:1976kj}.

Equation \eqref{GCA-EFEgen}, although interesting, is still rather ambiguous as a theory. In particular, i) to avoid deviations from GR in vacuum, one must provide a mechanism that drives the coupling tensor $\chi_{\mu \nu}{^{\alpha \beta}}$ to the product of two Kronecker deltas $\delta{^\alpha}{_\mu}\delta{^\beta}{_\nu}$ (up to a factor of $8 \pi G$) in vacuum, and ii) one should be able to construct a variational principle that generates the gravitational field equations. A first objective of the present work is to construct such a theory. We find that there is, in fact, a common solution of both of these problems at the cost of a modification of Eq.~\eqref{GCA-EFEgen}.

In this article, we provide a fundamental motivation for Eq.~\eqref{GCA-EFEgen} in the framework of semiclassical gravity. This motivation is useful because it provides a natural interpretation for a key parameter as a vacuum energy in our final theory. We then construct a general class of actions that can generate an equation similar to \eqref{GCA-EFEgen} in which the coupling tensor $\chi_{\mu \nu}{^{\alpha \beta}}$ is a function of a rank-2 tensor $A{_\mu}{^\alpha}$. These actions contain no derivatives of $A{_\mu}{^\alpha}$, i.e., this field is {\it nondynamical} (or auxiliary) and their variation leads to (algebraic) equations which constrain $A{_\mu}{^\alpha}$ to be a Kronecker delta in a vacuum.

We should remark that the strategy of employing auxiliary fields in modified gravity theories is not new. In the literature, other theories characterized by a similar setting have been explored. In Ref. \cite{Panietal2013}, for example, it was shown, under some rather general assumptions, that the introduction of auxiliary fields in GR will generally introduce higher derivatives of the energy-momentum tensor in the field equations. One of the assumptions in their approach is that the matter fields couple to the metric in the usual way, so that the matter Lagrangian reduces to the usual one in a local frame. In this article, we demonstrate that one can avoid higher derivatives of the energy-momentum tensor by relaxing this condition. In doing so, we obtain an example of a theory which generalizes the coupling between matter and gravity without introducing dynamical degrees of freedom or introducing higher derivatives of the matter fields.

We will study in detail an explicit example of such theories [the Minimal Exponential Measure (MEMe) model], and examine its basic features and phenomenology. Remarkably, we will find that for a single perfect fluid, the nondynamical auxiliary fields in the MEMe model induce a vector disformal transformation \cite{Bekenstein1993,Zuma2014,*Kimura2017,*Papadopoulos2018,*Domenech2018} of the metric within a matter distribution. While disformal generalized matter couplings have been explored in the recent literature \cite{BeltranJimenez2016,Gumrukcuoglu2019a,*Gumrukcuoglu2019b,*DeFelice2019}, we are not aware of any
disformal theory \cite{Bekenstein1993,Zuma2014,*Kimura2017,*Papadopoulos2018,*Domenech2018,BeltranJimenez2016,Gumrukcuoglu2019a,*Gumrukcuoglu2019b,*DeFelice2019, Kaloper2004,*Bettoni2013,*Zumalacarregui2013,*Deruelle2014,*Minamitsuji2014,*Watanabe2015,*Motohashi2016,*Domenech2015a,*Domenech2015b,*Sakstein2015,*Fujita2016,*vandeBruck2017,*Sato2018,*Firouzjahi2018,ClaytonMoffat1999,*ClaytonMoffat2000,*ClaytonMoffat2003,*Moffat2003,*Magueijo2009,*Moffat2016} which avoids introducing degrees of freedom through the use of auxiliary fields. A consequence of this is that gravitational waves propagate at the speed of light in a vacuum (consistent with the vacuum phenomenology of GR), but propagate with a different speed within a matter distribution. Additionally, we will show that MEMe cosmologies possess an unstable (de Sitter) inflationary era and also a (de Sitter) dark energy era in which the expansion rate is different. In fact, when a cosmological constant is introduced, the presence of the coupling tensor is able to alleviate, albeit not completely solve, the coincidence problem.

The paper is organized in the following way. Section II concerns a semiclassical gravity interpretation for \eqref{GCA-EFEgen}, which serves as a motivation for our work. Section III explores the general features of the rank-2 theory, in particular its derivation from a variational principle, the classification of different subclasses of theories, and the form of the field equations. In Sec. IV, the simplifications to the theory that follow when the matter model is a perfect fluid are described. Section V presents the MEMe model, its exact solution in the case of a single perfect fluid, and its general features. Section VI shows how data from gravitational wave signals can constrain the parameters of generalized coupling theories, and presents a parameter constraint for the MEMe model. Section VII contains the analysis of the cosmology of the MEMe model via phase space analysis. Section VIII concludes with a summary and discussion of future work.

We adopt the MTW signature ``$(-,+,+,+)$'' \cite{MTW} and use natural units $c=1$, defining $\kappa=8 \pi G$. Since index placement is critical in our analysis, the placement of indices in indexed quantities which appear as arguments in functions and and functionals will be indicated by dots.


%
%
\section{Semiclassical gravity framework}

Here, we propose a framework for semiclassical gravity which relaxes the coupling between matter and the gravitational field. The purpose of this section is to provide a fundamental motivation for generalized coupling theories; in particular, this discussion will allow us to later identify a key parameter in the theory with the vacuum energy. We first sketch a derivation of the semiclassical Einstein equations from the effective action. A more detailed discussion of these topics may be found in Refs. \cite{TomsSAP2012,WeinbergQFTII1996,ParkerToms2009QFTCS,BirrellDavies1984QFTCS,Visser2002}. We then discuss a modification of this derivation and obtain a framework in which the gravitational field does not couple directly to matter, but is mediated by a rank-4 tensor.

A quantum field theory for some field $\varphi$ on curved spacetime endowed with a classical metric $\mathfrak{g}^{\mu \nu}$ is defined by a generating functional $Z[J,\mathfrak{g}^{\cdot\cdot}]$, which has the formal functional integral expression
\begin{equation} \label{SCV-GeneratingFunctional}
Z[J,\mathfrak{g}^{\cdot\cdot}] = \int \mathcal{D}\varphi \, e^{i (S[\varphi]+\langle J \varphi \rangle_x) },
\end{equation}

\noindent where $\langle X \rangle_x:=\int X \, \sqrt{|\mathfrak{g}|} d^4 x$, $J$ is an external current,\footnote{The external current $J$ is typically introduced as a calculational tool for computing $N$-point correlation functions in quantum field theory and is set to zero at the end of the calculation; for additional details, consult Ref. \cite{RamondQFT,*SchwartzQFT}.} and $S[\varphi]$ is the action for matter fields. For the rest of this section, we suppress the functional dependence on $\mathfrak{g}^{\mu \nu}$, and unless stated otherwise, $Z[J]$ and all functionals constructed from it are implicitly functionals of $\mathfrak{g}^{\mu \nu}$. One can construct the following actionlike functional $W[J]$:
\begin{equation} \label{SCV-EnergyFunctional}
W[J]:=-i \ln Z[J].
\end{equation}

\noindent From $W[J]$, one may obtain the expression for the formal expectation value $\phi = \langle \varphi \rangle$ of the field $\varphi$,
\begin{equation} \label{SCV-ExpectationValueField}
\phi(x)=\left.\frac{\delta W[J]}{\delta J(x)}\right|_{J=0}.
\end{equation}

\noindent The field equations governing $\phi$ are obtained from the effective action $\Gamma[\phi]$, which may be implicitly defined as a functional Legendre transformation of $W[j]$,
\begin{equation} \label{SCV-EffectiveAction}
\Gamma[\phi]=-\langle j \phi \rangle_x+W[j],
\end{equation}

\noindent where now $j$ is an external current defined by
\begin{equation} \label{SCV-Implicitj}
j(x):=\frac{\delta \Gamma[\phi]}{\delta \phi(x)}.
\end{equation}

\noindent At this point, one can see that in the absence of the external current $j$, the functional derivative vanishes, and one recovers the principle of stationary action for $\Gamma[\phi]$. To one-loop order, the effective action has the form
\begin{equation} \label{SCV-EffectiveActionOneLoop}
\Gamma[\phi]=S[\phi] + \hbar \, \Gamma^{(1)}[\phi] + \mathcal{O}(\hbar^2),
\end{equation}

\noindent where $S[\phi]$ is the classical action evaluated on the expectation value $\phi$ and $\Gamma^{(1)}[\phi]$ is a functional, the explicit expression for which may be found in Ref. \cite{TomsSAP2012}. One may therefore interpret the effective action $\Gamma[\phi]$ to be a quantum corrected classical action. However, such an action is divergent, and, as is customary in quantum field theory, one typically adds counterterms in the Lagrangian to absorb these divergences; i.e., we perform a renormalization.

In curved spacetime, one can show that some of the divergent terms in $\Gamma[\phi]$ are purely geometrical; our analysis here focuses primarily on these terms. Therefore, an appropriate regularization at one loop level can be obtained by adding geometric counterterms to the effective action \cite{Sakharov1967,BirrellDavies1984QFTCS,ParkerToms2009QFTCS}. In particular, these counterterms will have the form
\begin{equation} \label{SCV-EffectiveActionCurvatureTerms}
\begin{aligned}
S_{ct}[\mathfrak{g}^{\cdot\cdot}]=\int d^4 x\sqrt{-\mathfrak{g}} \biggl[&\gamma_0 + \gamma_1 \, \mathfrak{R} + \gamma_{2,1} \, \mathfrak{R}^2 \\
&+ \gamma_{3,1} \, \mathfrak{C}^2 + \mathcal{O}(\mathfrak{R}{^\cdot}_{\cdot\cdot\cdot}{^3}) \biggr],
\end{aligned}
\end{equation}

\noindent where $\gamma_i$ are coupling constants, $\mathfrak{R}$ is the Ricci scalar, $\mathfrak{C}^2:=\mathfrak{C}_{\alpha \beta \mu \nu} \, \mathfrak{C}^{\alpha \beta \mu \nu}$ is the square of the Weyl tensor, and the remaining terms quadratic in curvature have been absorbed into the topological Gauss-Bonnet integral. The total action is therefore
\begin{equation} \label{SCV-SigmaActionSG}
\Sigma_{sg}[\phi,\mathfrak{g}^{\cdot\cdot}] = \Gamma_r[\phi,\mathfrak{g}^{\cdot\cdot}]+S_{ct}[\mathfrak{g}^{\cdot\cdot}].
\end{equation}
where $\Gamma_r$ includes $\phi$-dependent counterterms. At this point, one may recover the semiclassical Einstein action by choosing the constants $\gamma_i$ so that $S_{ct}[\mathfrak{g}^{\cdot\cdot}]$ completely cancels all curvature terms except for the Einstein Hilbert term and the vacuum energy term. Then, upon applying the stationary action principle to $\Sigma_{sg}[\phi,\mathfrak{g}^{\cdot\cdot}]$, one obtains
\begin{equation} \label{SCV-SCEFESigma}
\mathfrak{G}_{\mu \nu}+\Lambda \, \mathfrak{g}_{\mu \nu}=\kappa \, \mathfrak{T}_{\mu \nu}[\phi],
\end{equation}

\noindent where $\mathfrak{G}_{\mu \nu}$ is the Einstein tensor for the metric $\mathfrak{g}_{\mu\nu}$ and the energy-momentum tensor $\mathfrak{T}_{\mu \nu}[\phi]$ depends on the renormalized coupling constants, the expectation value of the field $\phi$, and $\mathfrak{g}_{\mu \nu}$.

Up to this point, the derivation we have presented is standard \cite{BirrellDavies1984QFTCS}. We now discuss a similar procedure which differs in that one drops the assumption that the metric that appears in the effective action is the gravitational metric. Instead, we postulate that the metric $\mathfrak{g}_{\mu \nu}$ is related to the gravitational metric $g_{\mu \nu}$ in the following way,
\begin{equation}\label{GCA-AuxiliarymetricSC}
\mathfrak{g}_{\mu \nu} := \chi{_{\mu \nu}}{^{\alpha \beta}} \, g_{\alpha \beta} ,
\end{equation}

\noindent where the rank-4 tensor $\chi{_{\mu \nu}}{^{\alpha \beta}}$ is constructed from other fields, which we shall specify later in this paper. We then propose a choice of constants $\gamma_i$ in the counterterm action $S_{ct}[\mathfrak{g}^{\cdot\cdot}]$ such that \textit{all} terms involving the curvature $\mathfrak{R}{^\alpha}_{\beta \mu \nu}$ in the effective action are canceled, including the Einstein-Hilbert term. In doing so, we effectively postulate that there is some mechanism which strongly suppresses the curvature terms in this model.

Assuming that the dynamics for coupling tensor $\chi{_{\mu \nu}}{^{\alpha \beta}}$ is provided by an action of the form $S_\chi[g^{\cdot\cdot},\chi{_{\cdot\cdot}}{^{\cdot\cdot}}]$,
the action for the renormalized one loop theory has the form
\begin{equation} \label{SCV-SigmaActionNew}
\Sigma[\phi,g^{\cdot\cdot},\chi{_{\cdot\cdot}}{^{\cdot\cdot}}] = \Gamma_r[\phi,\mathfrak{g}^{\cdot\cdot}]+S_{ct}[\mathfrak{g}^{\cdot\cdot}]+S_{G}[g^{\cdot\cdot}] +S_\chi[g^{\cdot\cdot},\chi{_{\cdot\cdot}}{^{\cdot\cdot}}],
\end{equation}
where $S_{G}[g^{\cdot\cdot}]$ is the Einstein-Hilbert action.

The action $\Sigma[\phi,g^{\cdot\cdot},\chi{_{\cdot\cdot}}{^{\cdot\cdot}}]$ now describes a framework in which the metric tensor $g_{\mu \nu}$ is no longer directly coupled to the matter fields; the coupling is mediated by the tensor $\chi{_{\mu \nu}}{^{\alpha \beta}}$. In the remainder of this article, we show that the tensor $\chi{_{\mu \nu}}{^{\alpha \beta}}$ does not necessarily require the introduction of additional dynamical degrees of freedom in the low-energy classical limit 
and that one can construct $\chi{_{\mu \nu}}{^{\alpha \beta}}$ entirely from nondynamical auxiliary fields. Also, since these auxiliary fields do not introduce derivatives of the energy-momentum tensor in the field equations, generalized coupling theories evade the no-go result of Ref. \cite{Panietal2013}. In fact, such a no-go result assumes that the matter fields couple to $g_{\mu \nu}$ in the usual way, which is no longer the case when the couplings between the matter fields and $g_{\mu \nu}$ are mediated by the tensor $\chi{_{\mu \nu}}{^{\alpha \beta}}$. We later identify and study a theory that is natural in the sense that $S_\chi$ is simply the vacuum energy term.


%
%
\section{Coupling tensor theories: general considerations}
\subsection{Couplings}
Here, we explore a class of theories in which the rank-4 coupling tensor $\chi{_{\mu \nu}}{^{\alpha \beta}}$ is constructed from invertible rank-2 coupling tensors $A{_\mu}{^\alpha}$, with inverse $\bar{A}{^\mu}{_\alpha}$. In particular, we assume that $\chi{_{\mu \nu}}{^{\alpha \beta}}$ may be decomposed in the following manner,
\begin{equation}\label{GCA-CouplingTensor}
\chi{_{\mu \nu}}{^{\alpha \beta}} = \Psi(A{_\cdot}{^\cdot}) \,  A{_\mu}{^\alpha} \, A{_\nu}{^\beta},
\end{equation}

\noindent where $\Psi(A{_\cdot}{^\cdot})$ is a scalar function of $A{_\mu}{^\alpha}$ that has the property $\Psi(\delta{_\cdot}{^\cdot})=1$ when $A{_\mu}{^\alpha}=\delta{_\mu}{^\alpha}$, where $\delta{_\mu}{^\alpha}$ is the Kronecker delta. Note that $A{_\mu}{^\alpha}=\delta{_\mu}{^\alpha}$ is a tensorial equation, but only when one index is raised and the other is lowered---this is because $\delta{_\mu}{^\alpha}$ is a tensor,\footnote{To see this, recall the expression $\delta{_\nu}{^\mu} = g^{\mu \sigma} \, g_{\nu \sigma}$.} but $\delta_{\mu \nu}$ and $\delta^{\mu \nu}$ are not. For this reason, it is important to pay particular attention to index placement when performing variations---see Appendix \ref{Appdx:Trace}. To simplify the analysis, the coupling tensors are assumed to be symmetric so that $A{_\mu}{^\alpha}=A{^\alpha}{_\mu}$ and $\bar{A}{^\mu}{_\alpha}=\bar{A}{_\alpha}{^\mu}$. From these tensors, one constructs a physical\footnote{In the sense that matter couples to $\mathfrak{g}_{\mu \nu}$.} metric $\mathfrak{g}_{\mu \nu}$ and its inverse $\mathfrak{g}^{\mu \nu}$,
\begin{equation}\label{GCA-Auxiliarymetric}
\mathfrak{g}_{\mu \nu} = \Psi(A{_\cdot}{^\cdot}) \,  A{_\mu}{^\alpha} \, A{_\nu}{^\beta} \, g_{\alpha \beta},
\end{equation}
\begin{equation}\label{GCA-AuxiliaryInvmetric}
\mathfrak{g}^{\alpha \beta} = \Psi^{-1}(A{_\cdot}{^\cdot}) \, \bar{A}{^\alpha}{_\mu} \, \bar{A}{^\beta}{_\nu} \, g^{\mu \nu} .
\end{equation}

\noindent Unless explicitly stated otherwise, indices are raised and lowered using the metric $g_{\mu \nu}$ and its inverse $g^{\mu \nu}$. The covariant derivative $\tilde{\nabla}_\mu$ is defined with respect to $\mathfrak{g}_{\mu \nu}$. We shall slightly abuse some terminology for the sake of convenience: throughout this article, we shall refer to the physical metric $\mathfrak{g}_{\mu \nu}$  as the ``Jordan frame'' metric and ${g}_{\mu \nu}$ as the ``Einstein frame'' metric.

Of course, since $A{_\mu}{^\alpha}$ are square matrices, Eq. (\ref{GCA-CouplingTensor}) does not describe the most general coupling that one can construct from $A{_\mu}{^\alpha}$; one could alternatively construct\footnote{The reader might observe that this is similar to what is done with tetrads $e_\mu{^a}$ in the tetrad formalism. The main difference here is that both indices of the tensor $\Theta{_\mu}{^\alpha}$ are in the coordinate basis (there are no internal Lorentz indices). However, one can nonetheless imagine $\Theta{_\mu}{^\alpha}$ to be a transformation of the metric tensor between one adapted to gravitational dynamics (the Einstein frame) and one adapted to matter (the Jordan frame).} $\chi{_{\mu \nu}}{^{\alpha \beta}}=\Theta{_\mu}{^\alpha} \, \Theta{_\nu}{^\beta}$ from a general power series in $A{_\mu}{^\alpha}$, labeled $\Theta{_\mu}{^\alpha}$, with coefficients that are scalar functions of $A{_\mu}{^\alpha}$. For instance, one may choose  $\Theta{_\mu}{^\alpha}=\exp(\delta{_\mu}{^\alpha}-A{_\mu}{^\alpha})$. It is also worth mentioning that one can also consider generalized couplings constructed from one-forms. For instance, one could consider a generalized vector disformal transformation of the form $\mathfrak{g}_{\mu \nu}=\omega^2 \, g_{\mu \nu} + \sigma \, A_\mu A_\nu$, where $\omega$ and $\sigma$ are functions of $A^2=g^{\mu \nu}A_\mu A_\nu$ and auxiliary scalar fields $\psi$, but one should be aware that unless $\omega$ and $\sigma$ are independent of $A_\mu$, such a coupling introduces an additional dependence on $g^{\mu \nu}$, which will generate additional terms in the gravitational equations. For the purposes of the present article, we will not explicitly\footnote{One might, however, imagine that the tensor $A{_\mu}{^\alpha}$ could in principle be a composite field constructed out of other auxiliary fields.} consider these alternative couplings, restricting only to those which have the form given in Eq. (\ref{GCA-CouplingTensor}).

The idea of considering theories of gravitation with two metrics related as in \eqref{GCA-Auxiliarymetric} or, more generally, in \eqref{GCA-AuxiliarymetricSC} offers an interesting connection with continuum mechanics and electromagnetism. Such relations have been studied before in various realizations; see Refs. \cite{BoehmerCarloni2018,*Pearson:2014iaa,*Boehmer:2014ipa,*Boehmer:2013ss,BeltranJimenez2016,Gumrukcuoglu2019a,*Gumrukcuoglu2019b,*DeFelice2019}. The difference with respect to these works is that in the present work, the behavior of the coupling tensor is explicitly given through a variational principle.

\subsection{Classification of theories}
We wish to construct theories with the property that the coupling tensors satisfy $A{_\mu}{^\alpha} = \delta{_\mu}{^\alpha}$ in the absence of matter. We do so by way of a variational principle, with a functional of the form
\begin{equation}\label{GCA-ActionAFunctional}
S_A[A{_\cdot}{^\cdot},g^{\cdot\cdot}]=- \frac{\lambda}{\kappa} \, \int d^4x \sqrt{-g} F(A{_\cdot}{^\cdot}).
\end{equation}

\noindent
There is no unique functional that yields $A{_\mu}{^\alpha} = \delta{_\mu}{^\alpha}$ as a solution. However, it is straightforward to construct such actions. A simple example is
\begin{equation}\label{GCA-ActionAPolySimp}
S_{A_q} = - \frac{\lambda}{ \,\kappa} \, \int d^4x \sqrt{-g} \, \left( \frac{1}{2} \, A - \frac{1}{4} A{_\beta}{^\alpha} \, A{_\alpha}{^\beta} \right).
\end{equation}

\noindent It is also straightforward to verify that the variation with respect to $A{_\alpha}{^\beta}$ yields the algebraic ``equation of motion'' $A{_\mu}{^\alpha} = \delta{_\mu}{^\alpha}$, as intended. More generally, one can construct a functional of the form
\begin{equation}\label{GCA-ActionAPoly}
S_{A_p} = - \frac{\lambda}{\kappa} \, \int d^4x \sqrt{-g} P(A{_\cdot}{^\cdot}),
\end{equation}

\noindent where $P(A{_\cdot}{^\cdot})$ is a polynomial function of $A{_\alpha}{^\beta}$ of finite order satisfying the property $P(\delta{_\cdot}{^\cdot})=1$. We call this class of theories \textit{polynomial class theories}. The coefficients for $P(A{_\cdot}{^\cdot})$ which yield the solution $A{_\mu}{^\alpha} = \delta{_\mu}{^\alpha}$ may be obtained by factoring the derivative of $P(A{_\cdot}{^\cdot})$ and demanding that at least one of the factors be $(A{_\mu}{^\alpha} - \delta{_\mu}{^\alpha})$. In particular, one can choose coefficients in the polynomial $P(A{_\cdot}{^\cdot})$ such that
\begin{equation}\label{GCA-FAPolynomialFactor}
\frac{\partial P}{\partial A{_\alpha}{^\beta}} =
(A{_\alpha}{^{\sigma}} - \delta{_\alpha}{^{\sigma}})f_{\sigma}^{\beta},
\end{equation}

\noindent where
$f_{\sigma}^{\beta}$ is some quotient polynomial.
It is not too difficult to demonstrate that to second order, the form of the action $S_{A_q}$ (\ref{GCA-ActionAPolySimp}) is the one that uniquely yields the solution $A{_\mu}{^\alpha} = \delta{_\mu}{^\alpha}$. One may note that higher-order polynomial class theories may yield additional nondegenerate solutions, but since $A{_\mu}{^\alpha}$ must satisfy an algebraic equation, it suffices to specify initial conditions that satisfy $A{_\mu}{^\alpha} = \delta{_\mu}{^\alpha}$.

Another class of simple theories have actions of the form
\begin{equation}\label{GCA-ActionADeterminant}
S_{A_e} = - \frac{\lambda}{\kappa} \, \int d^4x \sqrt{-g} \, |A|^n \, E(A{_\cdot}{^\cdot}),
\end{equation}

\noindent where $|A|=\det(A{_\cdot}{^\cdot})$, and again, $E(A{_\cdot}{^\cdot})$ is a function satisfying the property $E(\delta{_\cdot}{^\cdot})=1$. The variation of the above takes the form
\begin{equation}\label{GCA-ActionADeterminantVar}
\begin{aligned}
\delta S_{A_e} = &- \frac{\lambda}{\kappa} \, \int d^4x \sqrt{-g} \, |A|^n \times \\
& \left[ \left( \frac{\partial E}{\partial A{_\beta}{^\alpha}} + n \, E \, \bar{A}{^\beta}{_\alpha} \right)  \delta A{_\beta}{^\alpha} - \frac{1}{2} E \, g_{\mu \nu} \, \delta g^{\mu \nu} \right],
\end{aligned}
\end{equation}

\noindent and the variation with respect to $A{_\beta}{^\alpha}$ yields an equation which may (after a straightforward integration) be written as
\begin{equation}\label{GCA-ExponentialSolnB}
\frac{\partial \ln E}{\partial A{_\beta}{^\alpha}} = - n \, \bar{A}{^\beta}{_\alpha}.
\end{equation}

\noindent This suggests that $E(A{_\cdot}{^\cdot})$ has the form
\begin{equation}\label{GCA-ActionAExponentialIntegrand}
E(A{_\cdot}{^\cdot})=\exp\left(k-f_p(A{_\cdot}{^\cdot})\right),
\end{equation}

\noindent where $k=f_p(\delta{_\cdot}{^\cdot})$ and $f_p(A{_\cdot}{^\cdot})$ is a finite polynomial that, to second order and above, satisfies the following:
\begin{equation}\label{GCA-ActionAExponentialpolyIntegrand}
\frac{\partial (n \, A - f_p)}{\partial A{_\beta}{^\alpha}} =
(A{_\alpha}{^{\sigma}} - \delta{_\alpha}{^{\sigma}})f_{\sigma}^{\beta}.
\end{equation}

\noindent Again, $f_{\sigma}^{\beta}$ is some quotient polynomial. The simplest case is the choice $f_p = n \, A$ (in which case $k=4n$). Since Eq. (\ref{GCA-ActionAExponentialIntegrand}) is an exponential, theories of this type will be termed \textit{exponential class theories}.

The theories considered so far are \textit{homogeneous}, meaning that the actions depend explicitly on the tensor $A{_\mu}{^\alpha}$ or its inverse, but not both. One can also construct \textit{inhomogeneous} theories in which the action is an explicit functional of both $A{_\alpha}{^\beta}$ and $\bar{A}{^\alpha}{_\beta}$. It can be difficult to obtain analytical solutions for a general polynomial or exponential class theory, and it will become increasingly difficult to obtain analytical solutions for the more complicated inhomogeneous theories. For this reason, we will not study inhomogeneous theories any further, and will focus on the simplest theory which can be solved exactly for a perfect fluid.

\subsection{Gravitational action}
The theories described in the previous section are constructed so that when $A{_\mu}{^\alpha}=\delta{_\mu}{^\alpha}$, the action $S_A$ has the value
\begin{equation}\label{GCA-ActionAFunctionalOnShell}
S_A[\delta{_\cdot}{^\cdot},g^{\cdot\cdot}] = - \frac{\lambda \, V}{\kappa} ,
\end{equation}

\noindent $V := \int d^4x \sqrt{-g}$ being the four-volume. This is enforced by the requirement that when $A{_\mu}{^\alpha}=\delta{_\mu}{^\alpha}$, the integrand of the action $S_A$ satisfies the property $P(\delta{_\cdot}{^\cdot})=E(\delta{_\cdot}{^\cdot})=1$. Later, we find that the parameter $\lambda$ must be large in order to maintain consistency with late-time experimental and observational constraints, so we must add a counterterm $2 \lambda$ in the gravitational action. In particular, we assume that the dynamics for the gravitational metric $g_{\mu \nu}$ is provided by an action of the form
\begin{equation}\label{GCA-GravitationalAction}
\begin{aligned}
S_g[g^{\cdot\cdot}] &= \frac{1}{2 \, \kappa} \, \int d^4x \sqrt{-g} \left(R + 2 \, \tilde{\Lambda} \right) \\
&= \frac{1}{2 \, \kappa} \, \int d^4x \sqrt{-g} \left[R - 2 \, (\Lambda - \lambda) \right],
\end{aligned}
\end{equation}

\noindent where $\tilde{\Lambda}$ is a gravitational parameter related to the observed value of the cosmological constant $\Lambda$ according to the formula $\lambda-\tilde{\Lambda}=\Lambda$. It follows that in the generalized coupling theories we have constructed, $g_{\mu \nu}$ satisfies the vacuum Einstein field equations with cosmological constant $\Lambda=\lambda-\tilde{\Lambda}$,
\begin{equation}\label{GCA-EFEvac}
G_{\mu \nu} + \Lambda \, g_{\mu \nu} = 0 ,
\end{equation}

\noindent in the absence of matter. This result implies that general coupling theories do not avoid the fine-tuning problem associated with the cosmological constant, since one must require $|\lambda - \tilde{\Lambda}|/|\lambda| \ll 1$ to fit observational data. However, as we shall argue later, the fine-tuning problem can be mitigated to some degree in the particular generalized coupling theory we study.

\subsection{Generalized coupling in matter action}
Consider a matter action of the following form,
\begin{equation}\label{GCA-ActionMatter}
S_m= S_m [\phi, \mathfrak{g}^{\cdot \cdot}] = \int d^4x \sqrt{-\mathfrak{g}} \, L_m[\phi,\mathfrak{g}^{\cdot \cdot}] ,
\end{equation}

\noindent where $\phi$ (field indices suppressed) is a tensor field assumed to be minimally coupled to the metric $\mathfrak{g}^{\mu \nu}$. Up to boundary terms, the variation of the matter action $S_m$ has the following form,
\begin{equation}\label{GCA-ActionMatterVar}
\delta S_m = \int d^4x \sqrt{-\mathfrak{g}} \left(\mathbb{E}[\phi,\mathfrak{g}^{\cdot \cdot}] \, \delta \phi - \frac{1}{2} \, \mathfrak{T}_{\alpha \beta} \, \delta \mathfrak{g}^{\alpha \beta} \right),
\end{equation}

\noindent where $\mathbb{E}[\phi,\mathfrak{g}^{\cdot \cdot}]$ is the Euler-Lagrange operator yielding the field equations $\mathbb{E}[\phi,\mathfrak{g}^{\cdot \cdot}]=0$, and the Jordan frame energy-momentum tensor is defined as
\begin{equation}\label{GCA-EnergyMomTensorJordan}
\mathfrak{T}_{\alpha \beta} := - \frac{2}{\sqrt{-\mathfrak{g}}}\frac{\delta S_m}{\delta \mathfrak{g}^{\alpha \beta}}.
\end{equation}

\noindent One may relate $\mathfrak{T}_{\mu \nu}$ to the Einstein frame energy-momentum tensor $\tau_{\mu \nu}$ by making use of the chain rule
\begin{equation}\label{GCA-EnergyMomTensorEinstein}
\tau_{\mu \nu} := -\frac{2}{\sqrt{-g}}\frac{\delta S_m}{\delta g^{\mu \nu}} = - \frac{2 \, \Psi^2 \, |A|}{\sqrt{-\mathfrak{g}}}\frac{\delta S_m}{\delta \mathfrak{g}^{\alpha \beta}} \frac{\partial \mathfrak{g}^{\alpha \beta}}{\partial g^{\mu \nu}}.
\end{equation}

\noindent Using Eq. (\ref{GCA-AuxiliaryInvmetric}), one may obtain the following:
\begin{equation}\label{GCA-EnergyMomTensorEJ}
\tau_{\mu \nu} = \Psi \, |A| \, \mathfrak{T}_{\alpha \beta} \, \bar{A}{^\alpha}{_\mu} \, \bar{A}{^\beta}{_\nu} .
\end{equation}

\noindent The variation of $\mathfrak{g}^{\alpha \beta}$, as given by Eq. (\ref{GCA-AuxiliaryInvmetric}), is
\begin{equation}\label{GCA-AuxiliaryInvmetricVar}
\begin{aligned}
\delta \mathfrak{g}^{\alpha \beta}
%
%
%
%
    = & \Psi^{-1} \, \bar{A}{^\alpha}{_\mu} \, \bar{A}{^\beta}{_\nu} \, \delta g^{\mu \nu}\\
    & - \left(2 \, \mathfrak{g}^{\sigma (\beta} \, \bar{A}{^{\alpha)}}{_\tau} + \Psi^{-1} \, \mathfrak{g}^{\alpha \beta} \, \frac{\partial \Psi}{\partial A{_\sigma}{^\tau}} \right) \delta A{_\sigma}{^\tau} .
\end{aligned}
\end{equation}

\noindent The variation of the action then takes the following form,
\begin{equation}\label{GCA-ActionMatterVar2}
\begin{aligned}
\delta S_m
%
%
    = & \int d^4x \sqrt{-g} \biggl[\Psi^2 \, |A| \, \mathbb{E}[\phi,\mathfrak{g}^{\cdot \cdot}] \, \delta \phi - \frac{1}{2} \, \tau_{\mu \nu} \, \delta {g}^{\mu \nu} \\
    & + {\Psi^2 \, |A|}  \, \left(\mathfrak{T}_{\alpha \beta} \, \mathfrak{g}^{\sigma (\alpha} \, \bar{A}{^{\beta)}}{_\tau} + \frac{\mathfrak{T}}{2 \, \Psi} \, \frac{\partial \Psi}{\partial A{_\sigma}{^\tau}} \right) \delta A{_\sigma}{^\tau}\biggr]
.
\end{aligned}
\end{equation}

\noindent where for convenience we define the ``trace'' $\mathfrak{T}:=\mathfrak{T}_{\alpha \beta} \, \mathfrak{g}^{\alpha \beta}$. Note that, since the variation $\delta S_{m}$ depends on the variation $\delta A{_\sigma}{^\tau}$, the presence of matter will contribute additional terms to the field equation for $A{_\mu}{^\alpha}$, as we shall demonstrate shortly.

\subsection{General field equations}
We can now join together the previous results and give the general action for a generalized coupling theory. We have
\begin{equation}\label{GCA-GENAction}
\begin{aligned}
S[\phi,g^{\cdot\cdot},A{_{\cdot}}{^{\cdot}}]= \int d^4x \biggl\{&\left(R - 2 \, [\Lambda - \lambda(1-F)]\right)\sqrt{-{g}} \\
&+2 \, \kappa \, L_{m}[\phi,\mathfrak{g}^{\cdot\cdot}] \sqrt{-\mathfrak{g}} \biggr\},
\end{aligned}
\end{equation}
where $F=F(A{_\cdot}{^\cdot})$. Upon variation with respect to the metric and remembering that $A{_\mu}{^\alpha}$ is independent of $g_{\mu\nu}$, one obtains
\begin{equation}\label{GCA-GEN-GFE}
 G_{\mu \nu} +\left[  \Lambda- \, \lambda\left(1 - F\right)\right] \, g_{\mu \nu} = \kappa \, \Psi \, |A| \, \bar{A}{^\alpha}{_\mu} \, \bar{A}{^\beta}{_\nu} \, \mathfrak{T}_{\alpha \beta}.
\end{equation}
The form of this equation allows one to draw some general conclusions on the physics of these models. We notice immediately that the theory will generate a varying cosmological constant, which is dependent, via $A{_\mu}{^\alpha}$, on the matter distribution. Additionally, the presence of the quantity $\Psi \, |A| \, \bar{A}{^\alpha}{_\mu} \, \bar{A}{^\beta}{_\nu}$ contracted with $\mathfrak{T}_{\alpha \beta}$ ``scrambles'' the gravitational sources in a nontrivial way.

Notice also the differences between the \eqref{GCA-GEN-GFE} and \eqref{GCA-EFEgen}. In Eq. \eqref{GCA-EFEgen}, there is no effective cosmological term and the energy-momentum tensor is a function of the Einstein metric $g_{\mu\nu}$. This might suggest that the two theories are completely different. However, as stated in Ref. \citep{Carloni2017},  in \eqref{GCA-EFEgen},  $\chi_{\mu \nu}{^{\alpha \beta}}$ is completely general and thus can be chosen to return the structure of \eqref{GCA-GENAction}. In this sense, the two equations are still related.

The variation with respect to $A{_\mu}{^\alpha}$ yields the field equation for $A{_\mu}{^\alpha}$,
\begin{equation}\label{GCA-GEN-AFE}
(\delta{_\mu}{^{\alpha}}-A{_\mu}{^{\alpha}})f^{\nu}_{\alpha}={\Psi^2 \, |A|}  \, \left[\mathfrak{T}_{\alpha \beta} \, \mathfrak{g}^{\mu (\alpha} \, \bar{A}{^{\beta)}}{_\nu} + \mathfrak{T} \, \frac{1}{2 \, \Psi} \, \frac{\partial \Psi}{\partial A{_\mu}{^\nu}} \right],
\end{equation}
where, as before, $f^{\mu}_{\nu}$ is some tensor constructed from $A{_\mu}{^\alpha}$ such that
\begin{equation}
\frac{\delta F}{\delta A{_\mu}{^\nu}}=(A{_\nu}{^{\alpha}} - \delta{_\nu}{^{\alpha}})f^{\mu}_{\alpha}.
\end{equation}
As we have anticipated, the matter action $S_m$ contributes additional terms to the field equation (\ref{GCA-GEN-AFE}) for $A{_\mu}{^\alpha}$. These additional terms will generally drive the coupling tensor $A{_\mu}{^\alpha}$ away from the condition $A{_\mu}{^\alpha} = \delta{_\mu}{^\alpha}$. However, since we are assuming minimal coupling, Eq. (\ref{GCA-GEN-AFE}) contains no covariant derivatives of $A{_\mu}{^\alpha}$. Thus, the resulting field equation is an algebraic equation for $A{_\mu}{^\alpha}$, and the condition $A{_\mu}{^\alpha} = \delta{_\mu}{^\alpha}$ is only violated at points where $\mathfrak{T}_{\alpha \beta} \neq 0$. It follows that at points where the energy-momentum tensor $\mathfrak{T}_{\alpha \beta}$ vanishes, the coupling tensors satisfy $A{_\mu}{^\alpha} = \delta{_\mu}{^\alpha}$. In this sense, the field $A{_\mu}{^\alpha}$ has the same behavior as the torsion tensor in the context of Einstein-Cartan theory \citep{Hehl:1976kj}.

Before moving on to a specific example, we note that in general, the energy-momentum tensor $\mathfrak{T}_{\alpha \beta}$ contains factors of $\mathfrak{g}_{\mu \nu}$ and $\mathfrak{g}^{\alpha \beta}$, which generally introduce additional factors of $A{_\mu}{^\alpha}$ into the field equation. It follows that in general, Eq. \eqref{GCA-GEN-AFE} can be a high-order algebraic equation for $A{_\mu}{^\alpha}$. For example, in the context of a homogeneous polynomial class theory, a matter action like Eq. \eqref{GCA-ActionAPolySimp} will lead to an equation that is formally quadratic in ${A}{_\mu}{^\alpha}$. It is also possible that Eq. \eqref{GCA-GEN-AFE} may admit no real solutions. This implies that certain energy-momentum tensors $\mathfrak{T}_{\alpha \beta}$ may require that $A{_\mu}{^\alpha}$ be complex valued. For complex-valued $A{_\mu}{^\alpha}$, one generally has a complex metric $\mathfrak{g}_{\mu \nu}$ over a real manifold. The resulting complex matter action $S_m$ and the action $S_A$ should be replaced with the respective real actions $(S_m+S_m^*)/2$ and $(S_A+S_A^*)/2$ to ensure a consistent coupling to the real-valued gravitational degrees of freedom $g_{\mu \nu}$. Fortunately, this problem is not too severe; one can show that upon expanding $A{_\mu}{^\alpha}$ about $\delta{_\mu}{^\alpha}$, one does not need to consider complex values for $A{_\mu}{^\alpha}$ until fifth order in the expansion parameter. Also, the metric is only complex within a matter distribution; in a vacuum, one has $A{_\mu}{^\alpha}=\delta{_\mu}{^\alpha}$ so that $\mathfrak{g}_{\mu \nu}={g}_{\mu \nu}$. As we will show in the next section, it turns out that for a single perfect fluid, this problem does not arise in the specific theory we examine later in this article---we will in fact obtain an exact real-valued solution for $A{_\mu}{^\alpha}$.


%
%
\section{Perfect fluid ansatz}
Since we construct generalized coupling theories by way of a variational principle, it is appropriate to appeal to the variational principle for relativistic fluids, as discussed in Refs. \cite{Taub1954,Taub1969,Schutz1970,HawkingEllis,SchutzSorkin1977,Brown1993}. The variational principle for a relativistic perfect fluid may be formulated on an arbitrary background spacetime, which need not satisfy the Einstein equations, so that the relativistic Euler equations
\begin{equation}\label{EulerEq}
\mathfrak{g}^{\alpha \beta}\tilde{\nabla}_\beta \mathfrak{T}_{\alpha \gamma}=0
\end{equation}
are in general independent of the contracted Bianchi identities. We remind the reader that $\tilde{\nabla}_\mu$ is the covariant derivative compatible with the Jordan frame metric $\mathfrak{g}_{\mu \nu}$, and that matter is assumed to be minimally coupled to $\mathfrak{g}_{\mu \nu}$.

Equation \eqref{EulerEq} leads to another reason for appealing to a variational principle.  In generalized coupling theories, the contracted Bianchi identities do not imply that Eq. \eqref{EulerEq} is divergence free, but only that the source of the Einstein tensor satisfies the divergence-free property on shell; the relativistic Euler equations must be supplied by a variational principle. One can, on the other hand, show from the diffeomorphism invariance of the matter action that Eq. \eqref{EulerEq} must hold on shell. For a more detailed discussion, we refer the reader to Appendix \ref{Appdx:DivFree}.

Variational principles for perfect relativistic fluids usually involve constrained variations, and are formulated in terms of gradients of velocity potentials, so the covariant (lowered index) fluid four-velocity $u_\mu$ does not have a local dependence on the metric \cite{Schutz1970,Brown1993}. In terms of $u_\mu$, the energy-momentum tensor for a perfect fluid takes the form
\begin{equation} \label{GCA-EnergyMomentumPerfectFluid}
\mathfrak{T}_{\mu \nu} = \left(\rho + p\right)u_\mu u_\nu + p \> \mathfrak{g}_{\mu \nu}.
\end{equation}

\noindent It should be stressed that in general, $u_\mu \, u^\mu = u_\mu u_\nu g^{\mu \nu} \neq -1$; the fluid four-velocity only has unit norm with respect to the Jordan frame metric
\begin{equation} \label{GCA-UnitNorm}
u_\mu \, u_\nu \, \mathfrak{g}^{\mu \nu} = -1.
\end{equation}

\noindent Note that, since for a perfect fluid the trace is
\begin{equation} \label{GCA-EnergyMomentumPerfectFluidTrace}
\mathfrak{T}:=\mathfrak{T}_{\alpha \beta} \, \mathfrak{g}^{\alpha \beta}  = 3 p - \rho,
\end{equation}

\noindent only one factor of $\mathfrak{g}^{\alpha \beta}$ appears in the field equation \eqref{GCA-GEN-AFE}.

For perfect fluids, it is appropriate to employ the following ansatz for the solution,
\begin{equation}\label{GCA-Ansatz}
A{_\beta}{^\alpha} = {Y} \, \delta{_\beta}{^\alpha}  + {Z} \, u{_\beta} \, u{^\alpha} ,
\end{equation}

\noindent where $u_\mu$ is the fluid four-velocity. This ansatz will be justified in the next section for the specific model we study, but it can be considered nonetheless general for the case of a single perfect fluid. It is straightforward to show that the inverse of $A{_\mu}{^\alpha}$ as given in Eq. (\ref{GCA-Ansatz}) is (assuming $Y + \varepsilon \, Z \neq 0$)
\begin{equation}\label{GCA-AnsatzInv}
\bar{A}{^\alpha}{_\beta} = \frac{1}{Y}  \left(\delta{_\beta}{^\alpha} - \frac{Z}{Y + \varepsilon \, Z} \, u{^\alpha} \, u{_\beta} \right),
\end{equation}

\noindent where
\begin{equation}\label{GCA-unorm}
\varepsilon :=u^\mu \, u_\mu.
\end{equation}

\noindent Now one can define a timelike unit vector $U^\mu$ by rescaling $u^\mu$ (which is also assumed to be timelike, so that $\varepsilon<0$),
\begin{equation}\label{GCA-UnitFlowField}
U^\mu := {u^\mu}/{\sqrt{-\varepsilon}} ,
\end{equation}

\noindent and it follows that $u_\mu \, u_\nu = - \varepsilon U_\mu \, U_\nu$. Equation (\ref{GCA-Ansatz}) may then be written in the alternate form
\begin{equation}\label{GCA-AnsatzRS}
A{_\mu}{^\alpha} = {Y} \, \delta{_\mu}{^\alpha} - \varepsilon \, Z \, U{_\mu} \, U{^\alpha}.
\end{equation}

\noindent This form for $A{_\mu}{^\alpha}$ is useful for computing the determinant $|A|=\det(A{_\cdot}{^\cdot})$,
which reads
\begin{equation}\label{GCA-AnsatzRSDeterminant}
|A|=\det(A{_\cdot}{^\cdot})=Y^3 \left(Y + \varepsilon \, Z\right).
\end{equation}


%
%
\section{Minimal exponential measure model}\label{GCM}
Consider a matter action of the following form,
\begin{equation}\label{GCA-ActionMatterLambda}
S_{m_\lambda}= S_{m} - \frac{\lambda}{\kappa}\int d^4x \sqrt{-\mathfrak{g}},
\end{equation}

\noindent where $S_{m}$ is given by Eq. (\ref{GCA-ActionMatter}) and
\begin{equation}\label{det_g_goth}
\sqrt{-\mathfrak{g}}=\sqrt{-g}|A|\Psi^2.
\end{equation}
If the factor $\Psi^2$ is chosen such that the variation of the volume element $\sqrt{-\mathfrak{g}}$ yields the desired field equations for $A{_\mu}{^\alpha}$, then one may interpret the parameter $\lambda$ in Eq.~\eqref{GCA-ActionMatterLambda} to be something akin to a vacuum energy density generated by matter fields. This is in fact the interpretation provided by the semiclassical procedure outlined in Sec. II. In the framework of effective field theory, the value of $\lambda$ then corresponds to the energy scale at which a field theoretical description for matter breaks down.

To construct such a theory, we seek an expression for the factor $\Psi$ such that the variation of $\sqrt{-\mathfrak{g}}$ yields the field equation $A{_\mu}{^\alpha}=\delta{_\mu}{^\alpha}$ in a vacuum. Assuming $\Psi$ is an explicit function of $A{_\mu}{^\alpha}$ only, Eq. \eqref{det_g_goth} shows that the volume element in Eq.~\eqref{GCA-ActionMatterLambda}  has the same form of the integrand of the action $S_{A_e}$  for an exponential class theory in Eq. \eqref{GCA-ActionADeterminant}.

Hence, if we wish to recover the field equation $A{_\mu}{^\alpha}=\delta{_\mu}{^\alpha}$, $\Psi$ should have the form of an exponential of a polynomial in $A{_\mu}{^\alpha}$. The simplest such form for $\Psi$ is
\begin{equation}\label{GCA-PsiExpModel}
\Psi=\exp\left(\frac{4-A}{2 \, s}\right).
\end{equation}

\noindent This form for $\Psi$ introduces an exponential of the simplest polynomial of $A{_\mu}{^\alpha}$ into the Jordan frame measure $\int d^4x \sqrt{-\mathfrak{g}}$, and for this reason, we call the resulting theory the \textit{Minimal Exponential Measure} model, or the MEMe model. Defining the parameter
\begin{equation}\label{GCA-qparameter}
q:=\frac{\kappa}{\lambda},
\end{equation}

\noindent the variation of the volume functional is
\begin{equation}\label{GCA-ActionVolumeVarExpModel}
\begin{aligned}
- \frac{1}{q} \delta \int d^4x \sqrt{-\mathfrak{g}}
   =& - \frac{1}{q} \int d^4x \sqrt{-g}|A|\Psi^2 \times \\
   &\left[\left( \bar{A}{^\sigma}{_\tau} - \frac{1}{s}\, \delta{_\tau}{^\sigma} \right) \delta A{_\sigma}{^\tau} - \frac{1}{2} \, g_{\mu \nu} \, \delta g^{\mu \nu}\right]
,
\end{aligned}
\end{equation}

\noindent which leads to the parameter choice $s=1$. The variation of $S_m$ with respect to $A{_\sigma}{^\tau}$
\noindent 
yields the field equations
\begin{equation}\label{GCA-ExpFE}
\begin{aligned}
 \bar{A}{^\alpha}{_\beta} - \delta{_\beta}{^\alpha} = q \left[ \mathfrak{T}_{\mu \nu} \, \mathfrak{g}^{\alpha \nu} \, \bar{A}{^\mu}{_\beta} - (1/4) \mathfrak{T} \, \delta{_\beta}{^\alpha} \right] .
 \end{aligned}
\end{equation}

\noindent Upon multiplying through by ${A}{_\mu}{^\beta}$, we obtain an expression that is formally linear in the components of ${A}{_\mu}{^\alpha}$,
\begin{equation}\label{GCA-ExpFEs}
\begin{aligned}
{A}{_\beta}{^\alpha} - \delta{_\beta}{^\alpha} = q \left[ (1/4) \mathfrak{T} \, {A}{_\beta}{^\alpha} - \mathfrak{T}_{\beta \nu} \, \mathfrak{g}^{\alpha \nu} \right].
\end{aligned}
\end{equation}

\noindent We note here that the MEMe model is the simplest homogeneous theory that one can construct in the sense that one obtains an equation that is effectively linear in ${A}{_\mu}{^\alpha}$.

The trace of Eq. (\ref{GCA-ExpFEs}) yields an equation linear in the trace, $A=A_\alpha{^\alpha}$
\begin{equation}\label{GCA-ExpFEtr}
A-4 = q \, \mathfrak{T} \left( A/4 - 1 \right),
\end{equation}

\noindent which implies that $A=4$, which in turn implies $\Psi=1$, so that remarkably, all exponential factors disappear on shell. Note that in deriving this result, we have not yet made any assumption about the matter model; this result holds for any energy-momentum tensor.

We now consider the case of a perfect fluid. To justify the ansatz for $A{_\mu}{^\alpha}$ given in Eq. (\ref{GCA-Ansatz}), we consider the contraction of Eq. (\ref{GCA-ExpFEs}) with the vector $u_\alpha$. The result is
\begin{equation}\label{GCA-ExpFEsu}
 {A}{_\beta}{^\alpha} \, u_{\alpha} = - u_{\beta} \, \frac{1 + q \, \rho}{4 \, [4 - q \,(3 \, p - \rho)]},
\end{equation}

\noindent and it follows that $A{_\beta}{^\alpha} \, u_\alpha \propto u_\beta$, which is consistent with the ansatz for $A{_\mu}{^\alpha}$ given in Eq. (\ref{GCA-Ansatz}).

To obtain explicit expressions for $Y$ and $Z$, we use the ansatz (\ref{GCA-Ansatz}) to rewrite the field equation (\ref{GCA-ExpFEs}) in the form
\begin{equation}\label{GCA-ExpFieldEqsWform}
W{^\alpha}{_\beta} = W_1 \, \delta{_\beta}{^\alpha}+ W_2 \, u_\beta \, u^\alpha = 0,
\end{equation}

\noindent  where
\begin{equation}\label{GCA-ExpFieldEqsW}
\begin{aligned}
W_1 & = \frac{1}{4} Y \, [4 - q \, (3 \, p - \rho)]+p \, q - 1, \\
W_2 & = \frac{q \, (p + \rho)}{(\varepsilon \, Z + Y)^2}+\frac{1}{4} \, Z \, [4  - q \, (3 \, p - \rho)];
\end{aligned}
\end{equation}

\noindent i.e., Eq. (\ref{GCA-ExpFEs}) is reduced to the two scalar equations $W_1=0$ and $W_2=0$. As $W_1$ is completely independent of $\varepsilon$ and $Z$, it can be used to find an explicit expression for $Y$:
\begin{equation}\label{GCA-ExpYsoln}
Y = \frac{4 (1 - p \, q)}{4 - q \,  (3 \, p - \rho)}.
\end{equation}

\noindent The trace of the ansatz (\ref{GCA-Ansatz}) is $A = 4 Y + \varepsilon \, Z$, and since (\ref{GCA-ExpFEtr}) implies $A=4$, we obtain the following expression for $\varepsilon \, Z$,
\begin{equation}\label{GCA-ExpeZsoln}
\varepsilon \, Z = 4 \, (1-Y) = \frac{4 \, q \, (p + \rho)}{4 - q \,  (3 \, p - \rho)},
\end{equation}

\noindent where we have used Eq. (\ref{GCA-ExpYsoln}) for $Y$ in the second equality. We then solve the equation $W_2=0$ to obtain an expression for $Z$,
\begin{equation}\label{GCA-ExpZsoln}
Z = - \frac{q \, (p + \rho) [4 - q \, (3 \, p - \rho)]}{4 \, (q \, \rho + 1)^2},
\end{equation}

\noindent and from (\ref{GCA-ExpeZsoln}), we obtain the expression for $\varepsilon$,
\begin{equation}\label{GCA-Expepssoln}
\varepsilon = -\frac{16 \, (q \, \rho + 1)^2}{[4 - q \, (3 \, p - \rho)]^2}.
\end{equation}

Using Eq.~(\ref{GCA-ActionMatterLambda}) and Eq.~(\ref{GCA-GravitationalAction}) [cf. Eq. \eqref{GCA-GEN-GFE}], we obtain the gravitational field equations
\begin{equation}\label{GCA-ExpEFE}
\begin{aligned}
\resizebox{1.0\hsize}{!}{$
        G_{\mu \nu} + \left[\Lambda - \lambda \left(1 - e^{4-A} |A|\right)\right] g_{\mu \nu} = \kappa  e^{(4-A)/2} |A| \bar{A}{^\alpha}{_\mu} \bar{A}{^\beta}{_\nu} \mathfrak{T}_{\alpha \beta},
$}
\end{aligned}
\end{equation}

\noindent where the determinant $|A|$ is given by the explicit expression
\begin{equation}\label{GCA-ExpAdet}
|A|=\frac{256 \, (1 - p \, q)^3 (q \, \rho + 1)}{[4 - q \, (3 p - \rho) ]^4}.
\end{equation}

\noindent It is useful to write Eq.~\eqref{GCA-ExpEFE} in the form
\begin{equation}
G_{\mu \nu}=\kappa \, T_{\mu \nu},
\end{equation}
where $T_{\mu \nu}$ is the effective energy-momentum tensor defined by
\begin{equation}\label{GCA-ExpTmnEffDecomp}
T_{\mu \nu} = T_1 \, U_\mu \, U_\nu + T_2 \, g_{\mu \nu} ,
\end{equation}

\noindent and
\begin{equation}\label{GCA-ExpTmnEffDecompTs}
\begin{aligned}
T_1 & = |A| \, (p + \rho), \\
T_2 & = \frac{|A| \, (p \, q - 1) + 1}{q}-\frac{\Lambda}{\kappa}.
\end{aligned}
\end{equation}

\noindent Notice that, though $q$ appears in the denominator of the first term in $T_2$, $|A|$ is a function of $q$. Thus, in the limit $q \rightarrow 0$,
\begin{equation}
\begin{aligned}
T_1 & \rightarrow p + \rho, \\
T_2 & \rightarrow p-\Lambda/\kappa.
\end{aligned}
\end{equation}

\noindent In motivating the MEMe model, we have put forward the interpretation of $\lambda/\kappa=1/q$ as a form of vacuum energy density for quantum fields. If the MEMe model is regarded to be a low energy description for some quantum gravity theory, it is natural to expect the vacuum energy $1/q$ to be the Planck energy density, but whether this is the case ultimately depends on the specific details of the fundamental theory. Barring any specific knowledge about the underlying theory, we point out that $q$ can in principle be independent of the Planck scale. In fact, it turns out that the parameter $q$ sets the scale at which $\mathfrak{g}_{\mu \nu}$ fails to be a suitable spacetime metric and thus provides a natural regularization scale for quantum fields. To see this, note that for positive $q$, the determinant $|A|$ vanishes when $p \, q \rightarrow 1$, or when the pressure (assuming $w>0$) is on the order $1/q$. In this limit, $A{_\mu}{^\alpha}$ and $\mathfrak{g}_{\mu \nu}$ fail to be invertible, so that $\mathfrak{g}_{\mu \nu}$ cannot serve as a spacetime metric.  Thus, $q$ sets the scale at which the usual formulation of quantum field theory on a spacetime background given by $\mathfrak{g}_{\mu \nu}$ breaks down, and, as consequence, $1/q$ provides a natural regularization scale for the vacuum energy of quantum fields. This scale can in principle be independent of the Planck scale, as it concerns the breakdown in the effective spacetime metric $\mathfrak{g}_{\mu \nu}$, rather than the gravitational metric $g_{\mu \nu}$. In this sense, the MEMe model is phenomenologically compatible with a regularization scale for vacuum energy far below the Planck scale.

It is worth remarking that the breakdown of the description of the MEMe model as a bimetric theory does not imply that the model itself breaks down. Indeed, in the limit $p \, q \rightarrow 1$, the gravitational metric $g_{\mu \nu}$ remains well defined and the gravitational field equations have the form
\begin{equation}\label{GCA-EFELimit}
G_{\mu \nu} \approx \left(\lambda - \Lambda\right) g_{\mu \nu}.
\end{equation}

A similar mechanism is present in the case of negative $q$, which corresponds to a negative vacuum energy density for matter fields. For negative $q$, the determinant $|A|$ vanishes in the limit $\rho \, q \rightarrow -1$. In this case, one can also recover Eq. (\ref{GCA-EFELimit}), but now $\lambda$ is negative-valued. This suggests that for negative $\lambda$, one has a de Sitter phase in the limit $\rho \, q \rightarrow -1$, or when the density $\rho$ approaches the regularization scale $1/|q|$. We will revisit this in greater detail when we discuss the evolution of cosmologies obtained from the MEMe model.

The early de Sitter phase aside, there are more fundamental reasons to choose a negative value of $q$, or equivalently, a negative $\lambda$. One might suppose, for example, that in the fundamental theory, the fermionic contribution to the vacuum energy dominates, which would also imply a negative vacuum energy density and a negative value for $\lambda$. We remark that a negative vacuum energy density may be of particular interest in the context of string theory,\footnote{An important consideration, which falls beyond the scope of this article, is whether any theory which reduces to the MEMe model in some limit must lie in the ``swampland,'' in particular whether the MEMe model is incompatible with vacua allowed by string theory \cite{Palti2019,*Ooguri2019,*Garg2018,*Obied2018,*Vafa2005}.} in which a negative vacuum energy density is natural \cite{DanielssonVanRiet2018,*MaldacenaNunez2001}. A stronger case can me made if one imagines the MEMe model to be some limit of a dynamical theory; if $(\lambda/\kappa) \sqrt{-\mathfrak{g}}$ is interpreted to be potential for some dynamical theory, then dynamically stable solutions should lie at a minima of the potential. When evaluated on the solution $\bar{A}{^\sigma}{_\tau} = \delta{_\tau}{^\sigma}$, the second derivative of the potential taken with respect to $A{_\mu}{^\alpha}$, becomes
\begin{equation}\label{GCA-ActionVolumeVarExpModelSecond}
\left. \frac{\lambda}{\kappa}\frac{\partial^2 \sqrt{- \mathfrak{g}}}{\partial A{_\mu}{^\alpha} \> \partial A{_\nu}{^\beta}}\right|_{\bar{A}{^\sigma}{_\tau} = \delta{_\tau}{^\sigma}} = - \frac{\lambda \sqrt{-g}}{\kappa} \> \delta{^\mu}{_\beta} \> \delta{^\nu}{_\alpha} .
\end{equation}

\noindent The solution $\bar{A}{^\sigma}{_\tau} = \delta{_\tau}{^\sigma}$ corresponds to a minimum of the potential if $\lambda<0$, and the same solution is a maximum of the potential if $\lambda>0$. A negative $\lambda$ is therefore required to ensure the dynamical stability of the solutions, assuming that MEMe is a limit for some dynamical theory.

Earlier, we pointed out that generalized coupling models still suffer from the fine-tuning of the cosmological constant (in particular, $\Lambda/|\lambda|=|\lambda - \tilde{\Lambda}|/|\lambda| \ll 1$), and the MEMe model is no exception in this regard. However, we have argued that the vacuum energy in the MEMe model is independent of the Planck scale. If the magnitude of the vacuum energy is far below the Planck scale, the fine-tuning problem becomes less severe than that of the standard cosmological constant problem.


%
%
\section{Propagation speed for gravitational waves}\label{SecGW}
The MEMe model can be thought of as a bimetric theory, and unless the two metrics are related by a conformal transformation, bimetric theories generally have the property that the light cones defined by each metric do not necessarily coincide. To see this, we consider the form of the Jordan frame metric $\mathfrak{g}_{\mu \nu}$ for a perfect fluid given the ansatz (\ref{GCA-Ansatz}) for the tensor $A{_\mu}{^\alpha}$,
\begin{equation}\label{GCA-TransformedJordanMetric}
\mathfrak{g}_{\mu \nu} = \Psi \left[ Y^2 \, g_{\mu \nu} -\varepsilon Z \, (2 \, Y + \varepsilon \, Z) \, U_\mu \, U_\nu \right],
\end{equation}

\noindent which we note is a type of vector disformal transformation \cite{Zuma2014,*Kimura2017,*Papadopoulos2018,*Domenech2018}.

Since electromagnetic waves propagate according to the metric $\mathfrak{g}_{\mu \nu}$, and linearized gravitational waves propagate on a background metric $g_{\mu \nu}$, Eq. (\ref{GCA-TransformedJordanMetric}) may be used to compute the relative propagation speed between light and gravitational waves. The relationship between propagation speeds can be computed explicitly by considering a vector $k^\mu$ that is null with respect to the gravitational metric $g_{\mu \nu}$. In an orthonormal frame (defined with respect to $\mathfrak{g}_{\mu \nu}$) adapted to $u^\mu$, Eq. (\ref{GCA-TransformedJordanMetric}) yields the dispersion relation
\begin{equation}\label{GCA-Dispersion}
 - \left(1+\Psi \, Z \, (Y + \varepsilon \, Z)\right) (k^0)^2 + (k)^2=0.
\end{equation}

\noindent The speed of gravitational waves $c_g$ with respect to the metric $\mathfrak{g}_{\mu \nu}$ is then given by the expression
\begin{equation}\label{GCA-GWspeed}
c_g = c \, \sqrt{1+\Psi \, Z \, (Y + \varepsilon \, Z)},
\end{equation}

\noindent where $c$ is the speed of light. It is straightforward to show that if $q>0$, then for sufficiently small energy densities, $c_g<c$. We now define the quantity
\begin{equation}\label{GCA-GWspeeddiff}
\Delta := \left(c_g/c\right)^2 - 1 = \Psi \, Z \, (Y + \varepsilon \, Z).
\end{equation}

\noindent For the MEMe model, $\Delta$ has the expression
\begin{equation}\label{GCA-GWspeeddiffExp}
\Delta = - \frac{q \, \left(p + \rho\right)}{1 + \, q \, \rho} \approx - q \, \left(p + \rho\right)  + q^2 \, \rho \, \left(p + \rho\right)  + O(q^3) ,
\end{equation}

\noindent i.e., the theory predicts that gravitational waves propagating within a matter distribution will travel at a different speed compared to gravitational waves in a vacuum. It is worth pointing out that the difference in propagation speed is not necessarily a problem in the MEMe model. A generic argument is given in Ref. \cite{Babichev2008} and says that bimetric theories admitting two different speeds of light do not lead to causal paradoxes, so long as the system is well posed.

For positive $q$ or $\rho \, q < -1$, gravitational waves in the MEMe model slow down in the presence of matter ($\Delta<0$), so that they will refract in a manner similar to that of light in a medium. On the other hand, if $q$ is negative and $\rho < 1/|q|$, then gravitational waves can exceed the speed of light ($\Delta>0$). To lowest order in $q$, the amount by which gravitational waves speed up or slow down in a medium is linear in the energy density. The Earth provides a relatively dense medium through which detectable gravitational waves propagate,\footnote{One can use the constraints on the speed of gravitational waves from GW170817 and GRB170817A \cite{LIGOVIRGO_2017} to place constraints on $q$ (see for instance Ref. \cite{Bakeretal2017,*Creminelli2017,*Ezquiaga2017}), but this actually places a much a weaker constraint on $|q|$ than the propagation of gravitational waves through the Earth because the average energy density of baryonic and dark matter ($\Lambda$ does not significantly affect our result here) in the Universe is $30$ orders of magnitude smaller than the average energy density of the Earth.} so one may constrain the parameter $q$ using the uncertainty in time delay for gravitational waves. Recall that for a pair of gravitational wave detectors, the order of the arrival and time delay for gravitational wave signals can localize the source to a circular ring in the sky. However, such a localization assumes that gravitational wave signals propagate at $c_g=c$; $c_g>c$ will tend to ``widen'' the predicted localization rings (more precisely, the enclosed solid angle increases), and $c_g<c$ will tend to ``narrow'' the predicted localization rings. If one has an optical counterpart, then the location of the optical source in the sky may be different from the apparent localization if $q \neq 0$.

The gravitational detection GW170817 and its optical counterpart GRB170817A may be used to place a constraint on $|q|$ \cite{LIGOVIRGO_2017}. The signals were localized in the southern sky. The signal first arrived at the Virgo detector, then at the LIGO-Livingston 22 ms afterwards, and finally at LIGO-Hanford, 3 ms after its appearance at LIGO-Livingston \cite{LIGOVIRGO_2017MMO}. This order of events, combined with the fact that the signal was localized in the southern sky, indicates that the GW signal passed through the Earth. There is no significant discrepancy between the localization of GW170817 and the position of the signal GRB170817A, so we may immediately place a rough constraint on $|q|$ from the uncertainty in the difference of arrival times between different gravitational detectors, which we estimate to be roughly $5\%$. From the uncertainty in the time delay, one expects the uncertainty in $c_g$ to also be roughly $5\%$, so that $|q| < 0.1/\rho_E$. On average, the energy density of the Earth is $\rho_E \sim 5 \times 10^{12} \> \text{J}/\text{m}^3$, which places the following upper bound on $|q|$:
\begin{equation}\label{GCA-GWspeedConstraint}
|q| < 2 \times 10^{-14} \> \text{m}^3/\text{J}.
\end{equation}

\noindent Fundamentally, $\lambda/\kappa=1/q$ in the MEMe model corresponds to the vacuum energy density for quantum fields, but phenomenologically, the value of $|q|$ sets the value of the energy density at which the model breaks down. The highest energy scale probed to date is the TeV scale, which from $l \sim \hbar c/E$, corresponds to a length scale $l \sim 2 \times 10^{-19} \text{m}$. In a similar manner, one may use this to compute an inverse energy density via the expression $|q|=l^4/\hbar c$, so that if one expects new physics at the TeV scale, then one might expect $|q|\sim 5 \times 10^{-50} \text{m}^3/\text{J}$, which is about 35 orders of magnitude smaller than the constraints from GW170817; Eq. \eqref{GCA-GWspeedConstraint} is still a rather weak constraint. Of course, our analysis here is preliminary, and a more careful analysis may provide stronger constraints on $\lambda$.


%
%
\section{Cosmology}
We now explore  the features of cosmological models built on the field equations \eqref{GCA-ExpEFE} of the MEMe model. We will accomplish this task by employing dynamical systems techniques. Dynamical systems tools have now been used for a long time to explore cosmological models in GR (see, e.g., Ref. \citep{WainwrightEllis} and references therein) and modified gravity (e.g., Ref. \citep{Abdelwahab:2007jp,*Alho:2016gzi,*Bonanno:2011yx,*Carloni:2004kp,*Carloni:2007br,*Carloni:2007br,*Carloni:2007eu,*Carloni:2009jc,*Carloni:2009jc,*Carloni:2013hna,*Carloni:2014pha,*Carloni:2015bua,*Carloni:2015jla,*Carloni:2015lsa,*Carloni:2017ucm,*Leach:2006br,*Rosa:2019ejh,*Carloni:2018yoz,*Bahamonde:2017ize}). We will perform here a quick phase space analysis with the aim of evaluating the potential of \eqref{GCA-ExpEFE} to provide a theoretical framework for an inflationary/dark Universe.

The first step in constructing the cosmological model is to select a class of  reference observers. This step, which is often made tacitly in GR, is crucial in the context of theories like the MEMe model. A choice which is akin to the selection of fundamental observers in GR is the ``Jordan velocity'' $u_\mu$ of the matter fluid. However, such a choice would pose a serious problem:  the field $u_\mu$ is only a valid velocity frame if $\varepsilon$ is different from zero, i.e., $q \rho \neq -1$.

We prefer to use frames which have no such limitations, so as to avoid spurious singularities. A convenient choice is a frame $\bar{U}_\mu$  which is parallel to $u_\mu$ so that it is still orthogonal to the three-surfaces of homogeneity described by $\mathfrak{h}_{\mu \nu}=\mathfrak{g}_{\mu \nu}+u_\mu \, u_\nu$ and therefore prevents any ``tilting'' effect \citep{King:1972td}. In this way, we can suppose that for a homogenous and isotropic fluid source, the metric of the spacetime in the frame specified by $\bar{U}_\mu$ has the Friedmann-Lema\^{\i}tre-Robertson-Walker form
\begin{equation}\label{FLRW}
ds^2=-dt^2+S^2\left(t\right)\left[\frac{dr^2}{1-kr^2}+r^2d\Omega^2\right],
\end{equation}
where $k=-1,0,1$ is the spatial curvature, $d\Omega^2$ the infinitesimal solid angle and $S$ is the scale factor.

In the $\bar{U}_\mu$ frame, the cosmological equations can be written as
\begin{align}
 3q\left(H^2 +\frac{k}{S^2}\right)=& \frac{256 \kappa (1-pq)^3 (q \rho +1)^2}{[4+q (\rho-3p)]^4}+q\Lambda -\kappa,\label{FriedEq}\\
6q\left(\dot{H}+H^2\right)=& -\frac{256\kappa (p q-1)^3 (q \rho +1) [2-q(\rho+3p)]}{ [4+q (\rho-3p)]^4} \nonumber \\
&+2 (q \Lambda-\kappa),\label{RayEq}
\end{align}
 where $H=\dot{S}/S$ and we have assumed a barotropic equation of state $p=w\rho$ for the fluid. There is, however, an important caveat: the equations above only correspond to the equations for the MEMe model if $u_\mu$  is well defined. We must therefore require that $|A|\neq0$ ($\varepsilon\neq0$) in our analysis. In the same way, either  $\mathfrak{g}^{\alpha \sigma}\tilde{\nabla}_\sigma \mathfrak{T}_{\alpha \beta}=0$  or $\nabla^\alpha T_{\alpha \beta}=0$  give the conservation law
\begin{equation}\label{ConsLaw}
\dot{\rho}=-\frac{3 H \rho  (w+1) \left[q^2 \rho ^2 w (3 w-1)+\rho  (q-7 q w)+4\right]}{q^2
   \rho ^2 w (3 w-1)-q \rho  \left(3 w^2+13 w+2\right)+4}.
\end{equation}
As in GR, the three equations above are redundant. The structure of the above equations shows that in the MEMe model, the gravitation of a perfect fluid is very different form the GR equivalent.

Let us now construct the phase space. Assuming $H>0$, we define the dimensionless variables
\begin{align}
 \chi= \frac{q H^2}{\kappa},\quad &\quad K=\frac{k}{H^2 S^2},\\
 \Omega =\frac{\kappa  \rho }{3 H^2},\quad &\quad L=\frac{\Lambda }{3 H^2},
\end{align}
choosing also a dimensionless time variable $\mathcal{N}=\log \left(S/S_0\right)$ and indicating with a prime the derivative with respect to $\mathcal{N}$. The cosmological equations may be rewritten as the autonomous system
\begin{widetext}
\begin{align}\label{DynSysRed}
\begin{split}
\chi'=&2 (L-1) \chi +\frac{256 (3 \chi  \Omega +1)(3 w \chi  \Omega -1)^3 [3 \chi \Omega (1+3 w
   )-2] }{3 [3(1-3 w) \chi  \Omega +4]^4}-\frac{2}{3},\\
\Omega'=&\frac{1}{3} \Omega  \left\{\frac{2}{\chi }+6(1- L)-\frac{256 (3 \chi  \Omega +1)(3 w \chi  \Omega -1)^3 [3 \chi \Omega (1+3 w
   )-2] }{3 [3(1-3 w) \chi  \Omega +4]^4}\right.\\
   &\left.-\frac{9 (w+1)
    (3 w \chi  \Omega -1)[3 (3 w-1) \chi  \Omega -4]}{4-3 \left(3 w^2+13 w+2\right) \chi  \Omega +9 w (3 w-1) \chi ^2 \Omega
   ^2}\right\},\\
L'=&\frac{1}{3} L
   \left\{\frac{2}{\chi }+6(1- L)-\frac{256 (3 \chi  \Omega +1)(3 w \chi  \Omega -1)^3 [3 \chi \Omega (1+3 w
   )-2] }{3 [3(1-3 w) \chi  \Omega +4]^4}\right\},
\end{split}
\end{align}
\end{widetext}
together with the constraint
\begin{equation}
L-K-\frac{256 (3 \chi  \Omega +1)^2 (3 w \chi  \Omega -1)^3}{3 \chi  [3(1-3 w) \chi  \Omega +4]^4}-\frac{1}{3 \chi } =1,
\end{equation}
given by Eq.~\eqref{FriedEq}.

Although not immediately clear from the form of the \eqref{DynSysRed}, the system admits three invariant submanifolds $\Omega=0$, $L=0$, and $\chi=0$. Therefore, a global attractor can only exist at the origin. Setting to zero the lhs of \eqref{DynSysRed}, we find two fixed points ($\mathcal{A}$ and $\mathcal{B}$) and two lines of fixed points ($\mathcal{L}_1$ and $\mathcal{L}_2$). Notice that in the case $w=0$, $\mathcal{L}_2$ becomes asymptotic.

The solutions associated to the fixed points can be found using the modified Raychaudhuri equation \eqref{RayEq} written in the form
\begin{equation}\label{RaySol}
\begin{aligned}
\frac{\dot{H}}{H^2}=&\frac{128 \left(3 \chi _* \Omega _*+1\right) \left(3 w \chi _* \Omega _*-1\right)^3
   \left[3 \chi _*\Omega _* \left(3 w +1\right)-2\right]}{3\chi _* \left[3(1-3 w)
   \chi _* \Omega _*+4\right]^4} \\
  &+ \chi_* -1+\frac{1}{3L_*},
\end{aligned}
\end{equation}
where the asterisks indicate the values of the dynamical variables at a fixed point.

Using \eqref{RaySol}, we obtain the result that $\mathcal{A}$ and $\mathcal{B}$ correspond, respectively, to a Milne and Friedmann solution, whereas lines $\mathcal{L}_1$ and $\mathcal{L}_2$ correspond to de Sitter solutions with time constants
\begin{equation}
\begin{aligned}
\gamma_1&= \sqrt{\frac{\Lambda}{3}},\\
\gamma_2&=\sqrt{\frac{\Lambda  q-\kappa}{3q}+\frac{256 \kappa  \left(q \rho _0+1\right)^2 \left(q \rho _0
   w-1\right)^3}{q[4+3 q^4 \rho^4 _0\left(1-3 w\right)^4]}},
   \end{aligned}
\end{equation}
respectively. The cosmology then presents two different exponential expansion phases. Notice that the points in line~$\mathcal{L}_2$ are related to the behavior of the system in the limit $pq\rightarrow 1$ mentioned at the end of Sec. \ref{GCM}. As we have written our equations in the frame $\bar{U}_\mu$, the boundary $\rho q=-1$ does not correspond to any visible feature of the phase space. Nonetheless, one should bear in mind that physical orbits only correspond to $\Omega \chi\neq-1$ (and indeed  $\Omega \chi<-1$).

Writing the conservation law \eqref{ConsLaw} in the same way as the modified Raychaudhuri equation, we can deduce the behavior of the matter energy density at a fixed point,
\begin{equation}\label{ConsLawSol}
\frac{\dot{\rho}}{\rho}=\frac{3 H (w+1) \left(1-3 w \chi _* \Omega _*\right) \left[3 \chi _* \Omega_*(3 w-1) -4\right]}{9 w (3 w-1) \chi _*^2
   \Omega _*^2-3 \left(3 w^2+13 w+2\right) \chi _* \Omega _*+4},
\end{equation}
where $H$ is given by the modified Raychaudhuri equation \eqref{RaySol}. Apart from $\mathcal{L}_2$, the behavior of matter is the usual one. At the fixed point $\mathcal{L}_2$, the corrections to GR return a peculiar behavior in the energy density. In particular the energy density remains constant even if the spacetime is expanding. This implies that orbits near to this point have a very slowly decreasing energy density.

The Hartmann-Grobmann theorem can be used to deduce the stability of the fixed subspaces. It turns out that the lines are always attractors and the fixed points are saddles. In particular, as point $\mathcal{A}$ (the origin) is a saddle, no global attractor is present in the phase space. Therefore, orbits will go toward the lines $\mathcal{L}_1$ and $\mathcal{L}_2$.

\begin{figure*}[htbp]
\includegraphics[width=0.79 \textwidth]{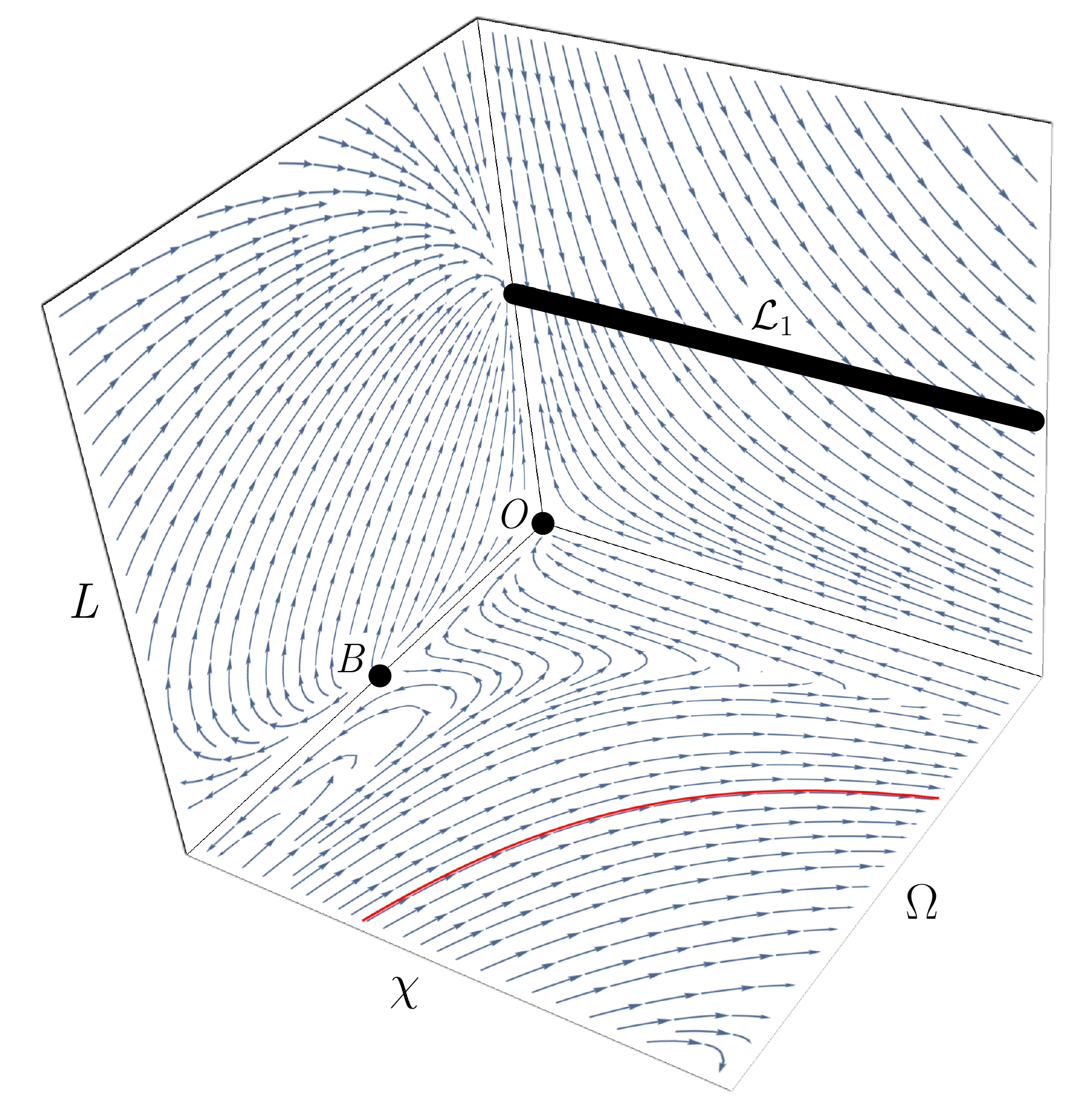}
\caption{A representation of the finite phase space of the system \eqref{DynSysRed} in the case of dust $w=0$ and $q<0$. The representation is constructed reporting on the $\chi<0$, $\Omega>0$, $L>0$ faces of the octant the invariant submanifolds $\chi=0$, $\Omega=0$, $L=0$. The black line of point corresponds to $\mathcal{L}_1$, whereas the red line represents the intersection of the singular plane $3\chi  \Omega +4=0$, which is present for $w=0$.}
\label{Fig:DS1}
\end{figure*}

\begin{figure*}[htbp]
\includegraphics[width=0.79 \textwidth]{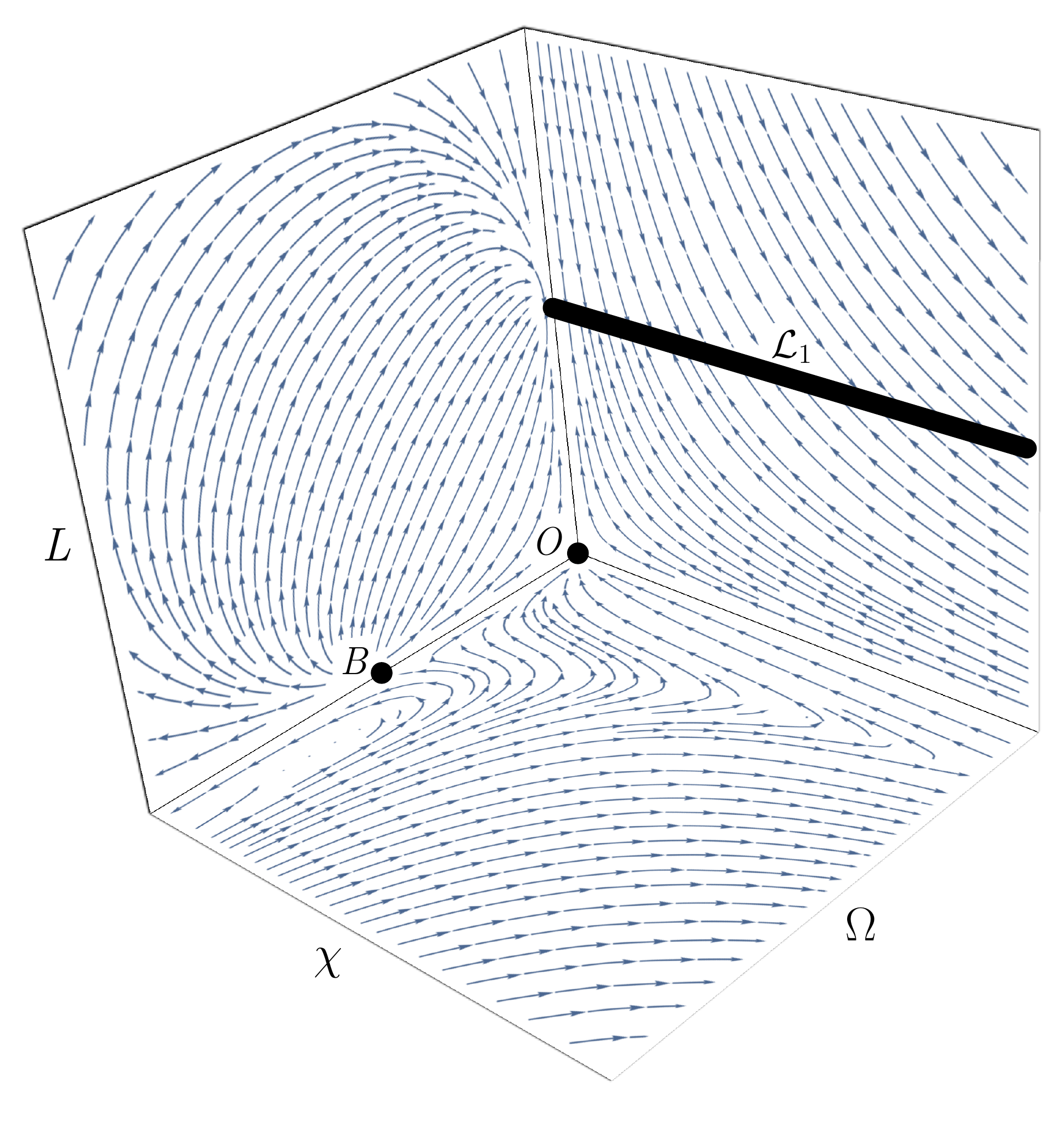}
\caption{A representation of the finite phase space of the system \eqref{DynSysRed} in the case of radiation $w=1/3$ and $q<0$. The representation is constructed reporting on the $\chi<0$, $\Omega>0$, $L>0$ faces of the octant the invariant submanifolds $\chi=0$, $\Omega=0$, $L=0$. The black line corresponds to line $\mathcal{L}_1$, and, differently form the dust case, there is no singular plane. This phase space is qualitatively very similar to the one in Fig. \ref{Fig:DS1}.}
\label{Fig:DS2}
\end{figure*}

The fixed points with their stability and the associated solutions for the scale factor and matter energy density are given in Table \ref{TableExp}. We illustrate the phase space in Figs. \ref{Fig:DS1} and \ref{Fig:DS2}.

As the phase space is not compact, one should study the behavior at infinity; however, for the purpose of this work, this analysis is not very relevant. This happens because, as we have mentioned, our equations cease to be valid if $|A|=0$, i.e., in the subspaces $3 w \Omega \chi= 1$ and $3\Omega \chi= -1$, respectively.  Consequently, all asymptotic fixed points that lie beyond the \textcolor{red}{l}ine~$\mathcal{L}_1$ and/or the boundary $3\Omega \chi= -1$ are irrelevant for our purposes. The same conclusion holds for the asymptotic points at $\{\Omega\rightarrow\infty,\chi\rightarrow 0 \}$ and $\{\Omega\rightarrow0,\chi\rightarrow\infty\}$: they do not constitute physically relevant states for the cosmology.

\begin{figure*}[htbp]
\includegraphics[width=0.75 \textwidth]{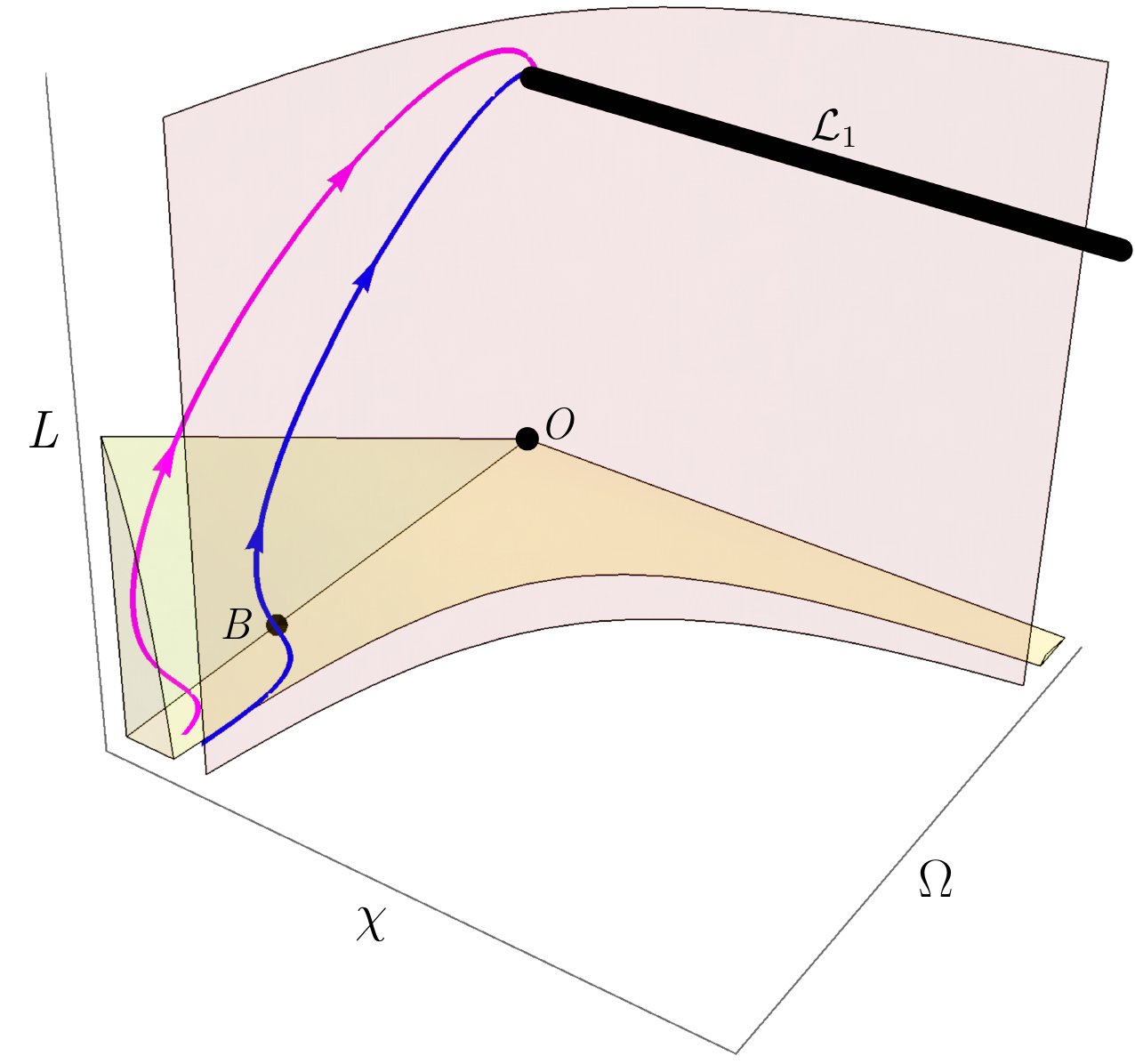}
\caption{Three different orbits in the case of dust $w=0$ and $q<0$. The blue and magenta orbits correspond to cosmic histories with a matter era close to that of GR. The yellow region corresponds to decelerating expansion, while the red surface delineates the $3\Omega \chi=-1$ ($q\rho=-1$) part of the phase space. For the sake of clarity here, we did not show the initial point of these orbits, as it is located far in the $\Omega$ direction, close to the surface $3\Omega \chi=-1$.}
\label{Fig:DS3}
\end{figure*}

As we know from Secs. \ref{GCM} and \ref{SecGW} that $q$ must be small, and if we wish to have a de Sitter phase at high densities, $q$ must also be negative. If we assume high densities at an early time, we expect that physically relevant initial conditions will be close to $3\Omega \chi= -1$ ($\rho \, q = -1$) at high $\Omega$. Therefore, one expects initial conditions for physical meaningful orbits to be close to the subspace $3\Omega \chi= -1$. Relevant orbits stemming from this subspace  ``bounce'' against the general Friedmann fixed point ($\mathcal{B}$) and then, maintaining a low value for $\chi$, the orbits evolve toward one of the attractors of $\mathcal{L}_1$. In Fig.~\ref{Fig:DS3} we give examples of these types of orbits.

These scenarios are interesting from the point of view of inflationary dark cosmology. In fact, they both admit an early and late accelerated expansion phase with different effective cosmological constants and thus solve naturally the ``graceful exit'' problem. The final value of the cosmological constant is $\Lambda$, while the inflationary expansion is given by a de Sitter solution with time constant very close to $\Lambda+\lambda$. The differences in the two histories are related to the expansion rate, the length, and time of occurrence of the decelerated expansion phase.

Another observation concerns the case in which the cosmological constant is zero. In this case, because of the existence of the invariant submanifold $L=0$, the dynamics is represented by the lower planes of Figs. \ref{Fig:DS1} and \ref{Fig:DS2}. It is clear that we still recover an inflationary phase, but no dark energy era is dynamically achievable as in the $L=0$ subspace, as the only finite attractor is point $\mathcal{A}$. In principle there are homoclinic orbits that start at $\{\Omega\rightarrow\infty,\chi\rightarrow 0 \}$ and bounce against point $\mathcal{B}$, but even neglecting the unphysical nature of the initial point, the picture in Fig. \ref{Fig:DS3} suggests that these orbits never represent accelerated expansion.

It is worth noticing that the scale factor $S$ we have deduced is not the one ``measured'' by the observers at rest with respect to matter, and therefore it is not in itself a physically meaningful quantity. Fortunately, for our choice of frame and symmetries, the behavior of the Jordan frame scale factor $S_J$ is easy to calculate. Using Eq. \eqref{GCA-TransformedJordanMetric}, we can easily show that
\begin{equation}
S_J= \Psi Y^2 S=\frac{16(1-p q)^2}{[4+q(\rho-3 p)]^2} S .
\end{equation}
We will use this relation to calculate $S_J$ corresponding to the solutions at the fixed points (see Table \ref{TableExp}). For negative $q$ and $0<w<1/3$, $S_J$ differs from $S$ by a factor of order unity as long as $\rho \leq 1/|q|$, so $S_J$ remains well defined when the determinant of the Jordan frame metric vanishes. On the other hand, for positive $q$ and $0<w<1/3$, $S_J$ vanishes as the pressure $p$ approaches $1/q$. Among the fixed points the biggest difference between $S$ and $S_J$ is present in the solutions associated to point $\mathcal{B}$, which only differ significantly at small $t$. These results suggest that critical constraints for the MEMe model can be found by looking closely at the phenomenology of the matter dominated era.

%
%
%
%
\begin{table*}[htp]
\caption{Fixed point, stability and associated solutions for the system \eqref{DynSysRed}.}
\begin{center}
\begin{tabular}{ccccccc}
\hline
Point & $\{\chi,\Omega,L, K\}$ & Attractor & Repeller & Scale factor & Energy density & Jordan scale factor \\
\hline\\
$\mathcal{A}$& $\left\{0,0,0,-1\right\}$ & Never & Never & $S= S_0 (t-t_0) $ & $\rho = 0$ & $S_J=S$\\ \\
$\mathcal{B}$ &$\left\{0,1,0,0\right\}$ & Never & Never & $S=S_0 (t-t_0)^{\frac{2}{3 (w+1)}} $ & $\rho = \frac{\rho _0} {(t-t_0)^{2}}$ & $S_J=\frac{16  \left[\left(t-t_0\right)^2-q \rho _0
   w\right]^2}{\left[q \rho _0 (1-3 w)+4 \left(t-t_0\right)^2\right]^2}S$ \\ \\ \hline\\
$\mathcal{L}_1$ &$\{\chi_0,0,1,0\}$ & $0\leq w\leq 1$ & Never & $S= S_0 e^{\gamma_1 (t-t_0)}$ & $\rho =0$ & $S_J=S$ \\ \\
$\mathcal{L}_2$ &$\{\frac{1}{3w\Omega_0},\Omega_0,1+w \Omega_0 ,0\}$ & $0< w\leq 1$ & Never & $S= S_0 e^{\gamma_2 (t-t_0)}$ & $\rho =\rho _0$ & $S_J=\frac{16 \left(1-q \rho _0 w\right){}^2}{\left(\text{q$\rho $}_0 (1-3 w)+4\right){}^2}S$\\ \\ \hline
  \end{tabular}
\end{center}
\label{TableExp}
\end{table*}%
%
%
%
%
%
%
%
%
%
%
%
%
%
%

%

%
%
\section{Conclusions}
In this article, we have proposed a new class of modified gravity theories in which the interaction of matter and spacetime is mediated by a rank-4 tensor $\chi{_{\mu \nu}}{^{\alpha \beta}} = \Psi(A{_\cdot}{^\cdot}) \,  A{_\mu}{^\alpha} \, A{_\nu}{^\beta}$, where $A{_\mu}{^\alpha}$ are nondynamical auxiliary fields. It is tempting to compare the role of this field with the one that the Higgs field has in particle physics: as the Higgs field determines the inertial mass of particles, $\chi_{\mu \nu}{^{\alpha \beta}}$ in a sense determines the active gravitational mass of matter. However, unlike the case of the Higgs mechanism, the fundamental mechanism that leads to a generalized coupling theory is not immediately evident at the present stage of investigation. Nonetheless, we have attempted to motivate generalized coupling theories in a framework that is more fundamental than simply specifying an action for the classical theory; in particular, we constructed the gravitational field equations via a procedure analogous to the one used to construct the semiclassical Einstein field equations. While this construction does require some fine-tuning, it provides a natural interpretation of $\lambda$ as the vacuum energy and may be useful as a starting point for finding a more fundamental theory from which a generalized coupling theory emerges as a limit.

It has been pointed out that auxiliary field theories can emerge in a strong coupling limit, as in the case of Ref. \cite{BanadosCohen2014}, and one might imagine that a generalized coupling theory also emerges in some limit. Alternately, one may imagine that the MEMe model emerges as the result of integrating out degrees of freedom in a more fundamental theory. However, even under such scenarios, fine-tuning problems remain: in order to obtain an effective action able to generate Eq.~\eqref{GCA-EFEgen}, we were forced to postulate the existence of some mechanism which suppresses the terms constructed from the curvature invariants for the metric $\mathfrak{g}_{\mu \nu}$. The need for such a mechanism may perhaps provide some guidance for constructing a more fundamental theory. In particular, a more detailed analysis is needed to determine whether a generalized coupling theory can naturally emerge from some effective field theory, or whether one must go beyond the framework of effective field theory to justify the suppression of curvature terms.

In light of the semiclassical derivation we have presented, it is also tempting to construct a fundamentally \textit{semiclassical} theory (in the sense that $g_{\mu \nu}$ is fundamentally classical) from the generalized coupling theories we have studied. A major obstacle to constructing such a theory concerns the fact that the time evolution for quantum fields depends on the background spacetime geometry, which is in turn dependent on the state of the quantum field. This interdependence will generically introduce nonlinear time evolution in quantum states \cite{Kibble1980,Carlip2008}. One way around this issue might be to employ some sort of measurement-feedback scheme for objective collapse models, which has been successfully implemented in Newtonian toy models \cite{Tilloy2019}, but a complete relativistic implementation of this approach is presently lacking. Even if a relativistic implementation can be constructed, one might expect the renormalized parameters in $\Sigma[\phi,g^{\cdot\cdot},\chi{_{\cdot\cdot}}{^{\cdot\cdot}}]$ to be scale dependent, so that the resulting energy-momentum tensor $\mathfrak{T}_{\mu \nu}$ is not unique; this nonuniqueness forms another obstacle to constructing a fundamental theory with our framework. Indeed, any attempt to construct a fundamental theory within our framework must supply a mechanism to regulate the divergences of quantum field theory. In the absence of such a mechanism, we take the conservative view that that our proposed theory is a low energy/coarse grained description for a more fundamental (and fully quantum) theory.

The generalized coupling theories we propose share features with other modified gravity theories. We have shown explicitly that a convenient framework of analysis is to interpret them as bimetric theories of gravity. Bimetric theories also modify the gravitational properties of standard matter in a manner akin to nonminimally coupled theories. Nonetheless, our generalized coupling theories have an advantage over these other frameworks in that the vacuum phenomenology remains essentially that of GR. Of course, whether one is truly in a vacuum depends on the details of the definition of matter in a gravitational theory. In GR, the source of the Einstein equations is modeled as a continuum, and in theories like the one we have proposed, one has to ask how well this approximation works in a given framework. Consider for example the issue of a photon traveling between two galaxies in a cosmological setting. Should one consider it as traveling in vacuum or in a continuum? In classical cosmology, these problems become almost irrelevant, but they might be of crucial importance in the generalized coupling case.

On a cosmological level, the field equations \eqref{GCA-GEN-GFE} show clearly the presence of a dynamical cosmological constant and a modification of the gravitational response to the thermodynamical potentials $\rho$ and $p$ for a fluid. The field equations \eqref{GCA-GEN-GFE} also share a common structure with many well-studied cosmological models like Loop Quantum Cosmology \citep{Ashtekar:2011ni}, the Randall-Sundrum type II braneworld model \citep{Randall:1999ee}, and some effective cosmological models, e.g., Refs. \citep{Copeland:2005xs,Lazkoz:2005bf}. Whether our theory can be related to any of those theories and how this can happen remains to be determined.

We have constructed a simple realization of our generalized coupling theories, which we call the Minimal Exponential Measure model, given by the action
\begin{equation}\label{GCA-MEMeAction}
\begin{aligned}
S[\phi,g^{\cdot\cdot},A{_{\cdot}}{^{\cdot}}]= & \int d^4x \biggl\{ \left[R - 2 \, \tilde{\Lambda} \right]\sqrt{-{g}}\\
&+2 \, \kappa\left(L_{m}[\phi,\mathfrak{g}^{\cdot\cdot}] - \frac{\lambda}{\kappa} \right) \sqrt{-\mathfrak{g}} \biggr\} ,
\end{aligned}
\end{equation}

\noindent where $\mathfrak{g}_{\mu \nu} = e^{(4-A)/2} \,  A{_\mu}{^\alpha} \, A{_\nu}{^\beta} \, g_{\alpha \beta}$ [see \eqref{GCA-GravitationalAction} and \eqref{GCA-ActionMatterLambda}], and $\tilde{\Lambda}=\Lambda - \lambda$. This choice of $\mathfrak{g}_{\mu \nu}$ is the simplest that one can make which does not require the choice of a particular form for the action of $A{_\mu}{^\alpha}$.

The MEMe model has three appealing features: i) it is equivalent to GR in a vacuum; ii) its (single) additional parameter $\lambda$ corresponds to a regularization scale that can in principle be independent of the Planck scale; and iii) for negative $q$ (with $\rho \, |q| < 1$), the MEMe model has inflationary behavior at early times without requiring additional dynamical degrees of freedom. Furthermore, a dynamical systems analysis indicates that the MEMe model can qualitatively describe cosmic histories which include an inflationary era, a graceful exit, and a dark energy era. There are a number of interesting questions that stem from these results. One might be concerned that the duration of the inflationary era is too dependent on the initial conditions. This is true if one thinks about inflation purely in terms of de Sitter expansion. Figure \ref{Fig:DS3} shows that there is a wide volume of the phase space which corresponds to accelerated inflation; therefore, a wide variety of parameter choices and initial conditions might lead to an inflationary phase of the right duration. The exact calculation, however, cannot be immediately made on the basis of the phase space analysis we have presented and will be left for future work. One might also ask whether, given the complex form of Eq. \eqref{ConsLaw}, the MEMe model is also able to predict the established sequence of cosmic eras, i.e., the fact that the Universe at early time is dominated by relativistic particles ($w=1/3$), successively by nonrelativistic matter ($w=0$), and then by spatial curvature. One can show that this is the case by recalling that $q$ must be small. Expanding in $q$ to first order, Eq.~\eqref{ConsLaw} reads
\begin{equation}
\dot{\rho}\approx-\frac{3}{4} H \rho  (w+1) \left(3 q \rho  (w+1)^2+4\right).
\end{equation}
Choosing the integration constant so that when $a=a_0$, $\rho=\rho_0$, one obtains
\begin{align}
\rho_d&=\frac{4  \rho _{d,0}a_0^3}{4 a^3+3 q \rho _{d,0} \left(a^3-a_0^3\right) } \\
\rho_r&=\frac{3  \rho _{r,0}a_0^4}{3 a^4+4 q \rho _{r,0}\left(a^4-a_0^4\right) }
\end{align}
for $w=0,1/3$, respectively. Since in our model we considered only $\rho q<-1$, $\rho_d$ and $\rho_r$ must always be finite, and we can neglect the singularity in the above expressions. For $\rho q<-1$, the behavior of both $\rho_d$ and $\rho_r$ follows closely the behavior of the standard cosmological model. This indicates that in MEMe cosmologies, the cosmic eras have the same chronological ordering as in the standard cosmological model.

It should be stressed, at this point, that the interesting cosmic histories we have found are only achievable through some (additional) fine-tuning. Indeed, we have to ensure that $\Lambda\ll|\lambda|$, but, as we have argued, this fine-tuning can be made less severe than that of the cosmological constant problem because, again, $\lambda$ can in principle be independent of the Planck scale.  A future research target will also be to determine whether the MEMe model can fit the observational data available so far. Such a task will entail a full analysis of the cosmological phenomenology to determine, for example, the inflationary power spectra for primordial fluctuations, the cosmic microwave background spectrum etc.

Since one can view the MEMe model (and indeed any generalized coupling theory) as a bimetric theory in the presence of matter, the MEMe model predicts a different propagation speed for gravitational waves within matter distributions. We were able to obtain a constraint on the parameter $q=\kappa/\lambda$ from the timing uncertainty of gravitational wave detections; however, this constraint is weak compared to the value of $q$ that one expects if new physics appears at the TeV scale. One can in principle obtain stronger constraints on $q$ by studying gravitational waves propagating through dense matter distributions, for instance the gravitational waves from the ringdown of a NS-BH merger propagating through the cloud of ejecta from the disrupted neutron star. Another interesting phenomenon predicted by the MEMe model is the refraction of gravitational waves by matter. These will also be the focus of future studies.


%
%


	\begin{acknowledgments}
	We thank Vitor Cardoso, David Hilditch, Michele Maggiore, Jos{\'e} Mimoso, Masato Minamitsuji, Shinji Mukohyama, Adrian del Rio Vega, and Daniele Vernieri for useful discussions. J.C.F. acknowledges support from FCT Grant No. PTDC/MAT-APL/30043/2017, and S.C. acknowledges support from FCT Grant No. UID/FIS/00099/2019 and Grant No. IF/00250/2013.
	\end{acknowledgments}


%
%



\appendix


%
%
\section{Variation of the trace} \label{Appdx:Trace}
In this Appendix, we examine the variation of the trace $A=A{_\sigma}{^\sigma}$. The purpose of this is to address a potential objection to the fact that we assume $A$ to be independent of $g^{\mu \nu}$. For instance, one might argue that, since $A=A_{\mu \nu} \, g^{\mu \nu}$, $A$ is dependent on $g^{\mu \nu}$ so that the variations will ultimately depend on $g^{\mu \nu}$.

The key point we wish to make here is that index placement matters when choosing the variables we vary, and that this choice determines whether the variations of $A$ depend on $g^{\mu \nu}$. To see this, consider the variation of $A=A_{\mu \nu} \, g^{\mu \nu}$,
\begin{equation}\label{GCA-APDX-TraceGDformVar}
\begin{aligned}
\delta A
	&= g^{\mu \nu} \, \delta A_{\mu \nu}  + A_{\mu \nu} \, \delta g^{\mu \nu} .
\end{aligned}
\end{equation}

\noindent If we regard $A_{\mu \nu}$ and $g^{\mu \nu}$ to be independent, the above expression suffices. However, if we instead demand that $A{_\mu}{^\nu}$ and $g^{\mu \nu}$ are independent variables, then we must rewrite $\delta A_{\mu \nu}$ in terms of $\delta A{_\mu}{^\nu}$ and $\delta g^{\mu \nu}$. In particular, we perform the variation of $A_{\mu \nu} = g_{\nu \sigma} \, A{_\mu}{^\sigma}:$
\begin{equation}\label{GCA-APDX-VarAlowered}
\begin{aligned}
\delta A_{\mu \nu}
	&= g_{\nu \sigma} \, \delta A{_\mu}{^\sigma} + A{_\mu}{^\sigma} \, \delta g_{\nu \sigma} \\
	&= g_{\nu \sigma} \, \delta A{_\mu}{^\sigma} - A{_\mu}{^\sigma} \, g_{\nu \alpha} \, g_{\beta \sigma} \, \delta g^{\alpha \beta}\\
	&= g_{\nu \sigma} \, \delta A{_\mu}{^\sigma} - A_{\mu \beta} \, g_{\nu \alpha} \, \delta g^{\alpha \beta}.
\end{aligned}
\end{equation}

\noindent The second line makes use of $\delta g_{\nu \sigma}=-g_{\nu \alpha} \, g_{\beta \sigma} \, \delta g^{\alpha \beta}$, which follows from the condition $\delta (g^{\mu \sigma} \, g_{\nu \sigma}) = 0$. Plugging this result back into (\ref{GCA-APDX-TraceGDformVar}) yields
\begin{equation}\label{GCA-APDX-TraceGDformVar2}
\begin{aligned}
\delta A
	&= g^{\mu \nu} \, g_{\nu \sigma} \, \delta A{_\mu}{^\sigma} - A_{\mu \beta} \, g^{\mu \nu} \, g_{\nu \alpha} \, \delta g^{\alpha \beta} + A_{\mu \nu} \, \delta g^{\mu \nu} \\
	&= g^{\mu \nu} \, g_{\nu \sigma} \, \delta A{_\mu}{^\sigma} - A_{\mu \nu} \, \delta g^{\mu \nu} + A_{\mu \nu} \, \delta g^{\mu \nu}.
\end{aligned}
\end{equation}

\noindent The last two terms cancel, and one obtains the result
\begin{equation}\label{GCA-APDX-TraceDeltaformVar}
\delta A = \delta{_\sigma}{^\mu} \, \delta A{_\mu}{^\sigma}.
\end{equation}

\noindent This demonstrates that $A$ is independent of $g^{\mu \nu}$ if we choose $A{_\mu}{^\nu}$ and $g^{\mu \nu}$ to be independent variables.


%
%
\section{Divergence-free property} \label{Appdx:DivFree}
Here, we show using variational methods that the source $T_{\mu \nu}$ of the Einstein tensor must satisfy $\nabla^\mu T_{\mu \nu}=0$ if the field equations are satisfied. We also demonstrate that the field equations also imply $\mathfrak{g}^{\alpha \sigma}\tilde{\nabla}_\sigma \mathfrak{T}_{\alpha \beta}=0$. Though the general proof is standard, and can be found in textbooks (Refs. \cite{Wald,*Weinberg}, for instance), we present it here to demonstrate that both $T_{\mu \nu}$ and $\mathfrak{T}_{\alpha \beta}$ must satisfy the divergence-free property on shell. Consider first an action of the form, defined on some domain $U$,
\begin{equation}\label{GCA-APDX-Action}
S[g^{\cdot\cdot},A{_\cdot}{^\cdot},\varphi] = S_{EH}[g^{\cdot\cdot}] + S_{A}[g^{\cdot\cdot},A{_\cdot}{^\cdot}] + S_{m}[g^{\cdot\cdot},A{_\cdot}{^\cdot},\varphi],
\end{equation}

\noindent where $S_{EH}
$ is the Einstein-Hilbert action. The variation has the form
\begin{equation}\label{GCA-APDX-ActionVar}
\begin{aligned}
\delta S = &\int_{U} d^4x \, \sqrt{-g} \biggl[\frac{1}{2 \, \kappa}(G_{\mu \nu} - \kappa \, T_{\mu \nu})\, \delta g^{\mu \nu} \\
&+ \frac{\Psi^2 \, |A|}{q} \, W{^\mu}{_\nu} \, \delta A{_\mu}{^\nu} + \Psi^2 \, |A| \, \mathbb{E}[\varphi,\mathfrak{g}^{\cdot \cdot}] \, \delta \varphi\biggr],
\end{aligned}
\end{equation}

\noindent where $S_A$ and $S_m$ are assumed to have a volume element of the form $d^4x \sqrt{-\mathfrak{g}}=d^4x \sqrt{-g} \Psi^2 |A|$, and
\begin{equation}\label{GCA-APDX-FieldEquationDefs}
\begin{aligned}
T_{\mu \nu} &:=\frac{-2}{\sqrt{-g}}\frac{\delta \left(S_A+S_m\right)}{\delta g^{\mu \nu}}, \\
W{^\alpha}{_\beta} &:= \frac{q}{\sqrt{-\mathfrak{g}}}\frac{\delta S}{\delta A{_\alpha}{^\beta}}, \\
\mathbb{E}[\varphi,\mathfrak{g}^{\cdot \cdot}] &:= \frac{1}{\sqrt{-\mathfrak{g}}}\frac{\delta S}{\delta \varphi}.
\end{aligned}
\end{equation}

\noindent Now, consider a differomorphism generated by a vector field $w^\mu$ which vanishes on the boundary $\partial U$,
\begin{equation}\label{GCA-APDX-DiffBC}
w^\mu|_{\partial U}=0.
\end{equation}

\noindent The action $S$, being constructed from covariant quantities, is diffeomorphism invariant; under the boundary condition (\ref{GCA-APDX-DiffBC}), a diffeomorphism generated by $w^\mu$ cannot change the value of the action, so that for an infinitesimal diffeomorphism of the form
\begin{equation}\label{GCA-APDX-Diff}
x\rightarrow x + \epsilon \, w^\mu,
\end{equation}

\noindent the first-order variation of the action resulting from the diffeomorphism must satisfy $\delta S=0$. If the infinitesimal parameter $\epsilon$ is constant, the variation of the metric takes the form:
\begin{equation}\label{GCA-APDX-Varmetric}
\delta_\epsilon g^{\mu \nu} = \epsilon \, \pounds_w g^{\mu \nu}
= 2 \, \epsilon \left(\nabla^{(\mu} w^{\nu)}\right).
\end{equation}

\noindent Upon integrating by parts and making use of the Bianchi identities $\nabla^\mu G_{\mu \nu} = 0$ and Eq. (\ref{GCA-APDX-DiffBC}), the condition $\delta S=0$ yields
\begin{equation}\label{GCA-APDX-ActionVarDiff}
\begin{aligned}
%
\epsilon \, \int_{U} d^4x \, \sqrt{-g} \, & w^\nu  \, \nabla^{\mu}T_{\mu \nu}
=-\int_{U} d^4x \, \sqrt{-g} \times \\
&\left[\frac{\Psi^2 \, |A|}{q} \, W{^\mu}{_\nu} \, \delta_\epsilon A{_\mu}{^\nu} + \Psi^2 \, |A| \, \mathbb{E}[\varphi,\mathfrak{g}^{\cdot \cdot}] \, \delta_\epsilon \varphi\right].
\end{aligned}
\end{equation}

\noindent If the field equations $\mathbb{E}[\varphi,\mathfrak{g}^{\cdot \cdot}]=0$ and $W{^\alpha}{_\beta}=0$ are satisfied, one has
\begin{equation}\label{GCA-APDX-ActionVarDiffOnshell}
\epsilon \, \int_{U} d^4x \, \sqrt{-g} \, w^\nu  \, \nabla^{\mu}T_{\mu \nu}=0,
\end{equation}

\noindent and if one demands that $\delta S=0$ for \textit{any} infinitesimal diffeomorphism, Eq. (\ref{GCA-APDX-ActionVarDiffOnshell}) must hold for all $w^\mu$, and it follows that $\nabla^{\mu}T_{\mu \nu}=0$. This demonstrates that $\nabla^{\mu}T_{\mu \nu}=0$ holds if the field equations $\mathbb{E}[\varphi,\mathfrak{g}^{\cdot \cdot}]=0$ and $W{^\alpha}{_\beta}=0$ are satisfied.

A similar argument may be used to demonstrate that $\mathfrak{T}_{\alpha \beta}$ satisfies the divergence-free property on shell. In particular, one may show that on shell, $\mathfrak{g}^{\alpha \sigma}\tilde{\nabla}_\sigma \mathfrak{T}_{\alpha \beta}=0$. Recalling that $S_m= S_m [\varphi, \mathfrak{g}^{\cdot \cdot}]$ [cf. (\ref{GCA-ActionMatter})], the variation of $S_m$ takes the form (\ref{GCA-ActionMatterVar})
\begin{equation}\label{GCA-APDX-ActionMatterVar}
\delta S_m = \int d^4x \sqrt{-\mathfrak{g}} \left(\mathbb{E}[\varphi,\mathfrak{g}^{\cdot \cdot}] \, \delta \varphi - \frac{1}{2} \, \mathfrak{T}_{\alpha \beta} \, \delta \mathfrak{g}^{\alpha \beta} \right),
\end{equation}

\noindent where $\mathfrak{T}_{\alpha \beta}$ is defined in Eq. (\ref{GCA-EnergyMomTensorJordan}). Under the diffeomorphism (\ref{GCA-APDX-Diff}), the variation in $\mathfrak{g}^{\alpha \beta}$ has the form
\begin{equation}\label{GCA-APDX-VarJmetric}
\delta_\epsilon \mathfrak{g}^{\alpha \beta} = \epsilon \, \pounds_w \mathfrak{g}^{\alpha \beta}
= 2 \, \epsilon \, \tilde{\nabla}_\sigma \left(\mathfrak{g}^{ \sigma (\alpha} w^{\beta)}\right),
\end{equation}

\noindent and upon demanding $\delta S_m = 0$, one obtains
\begin{equation}\label{GCA-APDX-ActionVarDiffJ}
\epsilon \int_{U} d^4x \sqrt{-g} w^\beta\mathfrak{g}^{\alpha \sigma}\tilde{\nabla}_\sigma\mathfrak{T}_{\alpha \beta}
=-\int d^4x \sqrt{-\mathfrak{g}} \left[\mathbb{E}[\varphi,\mathfrak{g}^{\cdot \cdot}]\delta_\epsilon \varphi\right].
\end{equation}

\noindent On shell, $\mathbb{E}[\varphi,\mathfrak{g}^{\cdot \cdot}]=0$, and upon demanding that $\delta S_m = 0$ for all $w^\mu$, $\mathfrak{g}^{\alpha \sigma}\tilde{\nabla}_\sigma \mathfrak{T}_{\alpha \beta}=0$. Note that this result depends only on the diffeomorphism invariance of $S_m$, and does not require that $\mathfrak{g}_{\mu \nu}$ satisfy the gravitational field equations---the argument is valid for any metric $\mathfrak{g}_{\mu \nu}$. This demonstrates that the property $\mathfrak{g}^{\alpha \sigma}\tilde{\nabla}_\sigma \mathfrak{T}_{\alpha \beta}=0$ is independent of the Bianchi identities. We note that both $\nabla^{\mu}T_{\mu \nu}=0$ and $\mathfrak{g}^{\alpha \sigma}\tilde{\nabla}_\sigma \mathfrak{T}_{\alpha \beta}=0$ require that the field equations $\mathbb{E}[\varphi,\mathfrak{g}^{\cdot \cdot}]=0$ are satisfied, and that $\nabla^{\mu}T_{\mu \nu}=0$ depends also on the equation $W{^\alpha}{_\beta}=0$. Furthermore, one can derive $\nabla^{\mu}T_{\mu \nu}=0$ \textit{without} including the Einstein-Hilbert action. Thus, the view that the divergence-free property of the energy-momentum tensor is enforced by the gravitational field equations is somewhat inaccurate from a fundamental perspective. A more accurate view is that the divergence-free property and local conservation laws follow from the diffeomorphism invariance of the action and the field equations.


%
%


    \bibliography{GenCoup}


\end{document}